\documentclass[letterpaper,twocolumn,reprint,amsmath,amssymb,pra]{revtex4-2}
\usepackage{mathptmx}

\usepackage[T1]{fontenc}
\usepackage[utf8]{inputenc}
\usepackage{color}
\usepackage{amsmath}
\usepackage{graphicx}
\usepackage[pdfusetitle,
 bookmarks=true,bookmarksnumbered=false,bookmarksopen=false,
 breaklinks=false,pdfborder={0 0 1},backref=false,colorlinks=true]
 {hyperref}
\hypersetup{
 pdfborderstyle=}

\makeatletter

\pdfpageheight\paperheight
\pdfpagewidth\paperwidth

\providecommand{\tabularnewline}{\\}

\makeatother

\begin{document}
\title{Broadening the reflection band of atomic mirrors by engineering the
emitter-waveguide coupling}
\author{C.-Y. Lee$^{1}$}
\email{lcy511085@gmail.com}

\author{G.-D. Lin$^{1,2,3}$}
\author{H. H. Jen$^{4,1,5}$}
\author{K.-T. Lin$^{2}$}
\email{alex968.tw@gmail.com}

\affiliation{$^{1}$Physics Division, National Center for Theoretical Sciences,
Taipei 10617, Taiwan~~~\linebreak{}
 $^{2}$Trapped-Ion Quantum Computing Laboratory, Hon Hai Research
Institute, Taipei 11492, Taiwan~~~\linebreak{}
 $^{3}$Department of Physics and Center for Quantum Science and Engineering,
National Taiwan University, Taipei 10617, Taiwan~~~\linebreak{}
 $^{4}$Institute of Atomic and Molecular Sciences, Academia Sinica,
Taipei 10617, Taiwan~~~\linebreak{}
 $^{5}$Department of Physics, National Central University, Taoyuan
City 320317, Taiwan}
\begin{abstract}
We theoretically investigate single-photon coherent transport through
a finite waveguide quantum electrodynamics (QED) array with spatially
modulated emitter-waveguide couplings. By combining exact scattering
calculations with spectral factorization and a semiclassical WKB analysis,
we identify how the coupling profile controls the reflection bandwidth
and its finite-size dependence. A Gaussian transverse displacement
produces an exponential-of-Gaussian decay-rate profile that broadens
the high-reflectivity window and suppresses the boundary-induced Fabry-Perot
oscillations compared with a uniform array. The transmission zeros
remain exactly degenerate at the atomic resonance for arbitrary positive
coupling profiles at non-Bragg spacings. In contrast, the collective
poles reorganize under the spatial modulation, redistributing their
radiative widths and preventing individual subradiant resonances from
terminating the reflection window. We further derive an algebraic
$1/N$ scaling law for the contraction of the bandwidth toward its
bulk value. The bulk bandwidth is determined by the largest local
decay rate, whereas the finite-size correction increases as the profile
becomes sharper, leading to an optimal profile width. These results
establish spatial coupling engineering as a practical tool for designing
broadband atomic mirrors in atom-nanofiber and circuit-QED platforms. 
\end{abstract}
\maketitle
\hypersetup{ colorlinks=true, urlcolor=blue, linkcolor=blue, citecolor=blue }

\section{Introduction}

Waveguide quantum electrodynamics (QED) addresses the coherent interaction
between quantum emitters and the guided one-dimensional (1D) photonic
continuum of a nanophotonic or microwave waveguide~\citep{Shen2005,Zhou2008,Roy2017,Chang2018,Sheremet2023}.
Emitters separated by many wavelengths remain coupled, and their radiative
properties become collective~\citep{Lalumiere2013,Solano2017}. This
long-range behavior arises from the guided mode, which mediates an
interaction whose range is set by an accumulated propagation phase
rather than by a decaying near field. This regime is now accessible
in several platforms, including superconducting qubits coupled to
open transmission lines~\citep{Astafiev2010,Hoi2011,vanLoo2013,Mirhosseini2019}
and cold atoms trapped in the evanescent field of tapered optical
nanofibers~\citep{Vetsch2010,Goban2012,Corzo2019} and photonic crystal
waveguides~\citep{Goban2014,Goban2015}.

One application of collective radiance is the atomic mirror. In a
periodic array whose spacing satisfies the Bragg condition, the single-emitter
reflection amplitudes add in phase, and the array reflects a resonant
guided photon with near-unity efficiency~\citep{Deutsch1995,Chang2012}.
Reflection of this kind has been observed with atoms trapped along
an optical nanofiber~\citep{Corzo2016,Sorensen2016}, with a structured
atomic monolayer in free space~\citep{Rui2020}, and with transmon
qubits acting as atom-like mirrors in circuit QED~\citep{Mirhosseini2019}.
Cooperative resonances in two-dimensional atomic arrays have also
been analyzed theoretically~\citep{Shahmoon2017}. Since such a mirror
is assembled from the same emitters that carry the quantum information,
it is attractive as an in-line element for photon routing and quantum
memory~\citep{Hoi2011,Corzo2019}.

The performance of a uniform array as a mirror is limited in two respects.
First, for a Bragg-spaced array, total reflection occurs only on atomic
resonance and degrades rapidly with detuning~\citep{Chang2012}.
Away from the Bragg condition, a finite high-reflectivity window opens,
but its width is fixed by the collective bandgap of the uniform array.
Second, a finite array terminates abruptly against the waveguide continuum,
and the impedance seen by a guided photon changes discontinuously
at the two ends. The resulting boundary reflections produce Fabry-Perot
ripple at the edges of the reflection band, as it does in any finite
Bragg stack~\citep{Yariv2003}.

Non-uniform and non-Bragg arrangements have been examined from several
directions: anti-Bragg spacings~\citep{Poddubny2022}; quasiperiodic
arrays, in which the single-photon scattering states localize~\citep{Zhang2026};
array geometries optimized for selective radiance and photon storage~\citep{AsenjoGarcia2017,Manzoni2018};
disorder-assisted localization and bound or subradiant excitations~\citep{Jen2020,Jen2021};
and transverse-displacement profiles used to control collective relaxation
and excitation localization~\citep{Lee2025}. In the context of atomic
mirrors, Ref.~\citep{Wang2025} broadens the ultrahigh-reflection
window by tuning the interatomic spacing toward the Bragg distance
and by imposing a gradient modulation of the atomic transition frequencies.
In contrast, we keep the spacing and the transition frequencies fixed
and modulate only the spatial profile of the emitter-waveguide coupling.
Such profiles have so far been used mainly to control collective decay
and excitation localization, and their role in single-photon transport
remains underexplored. In particular, to our knowledge, how a spatially
modulated coupling reshapes the collective poles of the scattering
matrix, and thereby the reflection bandwidth of a finite array, has
not been addressed.

In this work, we investigate single-photon transport through an emitter
array with a Gaussian transverse displacement, which generates an
exponential-of-Gaussian decay-rate profile while its sum remains fixed.
We combine exact scattering calculations with a factorization in terms
of transmission zeros and collective poles. We further develop a semiclassical
description for arrays that support a local band structure, which
gives the explicit dependence of the reflection bandwidth on the emitter
number and the coupling profile.

Our results show that coupling engineering broadens the reflection
window by redistributing the radiative widths of the collective modes.
The bulk bandwidth is determined by the largest local decay rate,
whereas the finite-size contraction is controlled by the number of
emitters within the strongly coupled region. The competition between
these two effects selects an optimal envelope width. These findings
provide insights into controllable single-photon transport and offer
practical design principles for broadband atomic mirrors. The required
coupling profiles can be implemented with atoms trapped near optical
nanofibers~\citep{Vetsch2010,Goban2012,Corzo2019} or with superconducting
qubits coupled to open transmission lines~\citep{Astafiev2010,Hoi2011,vanLoo2013,Mirhosseini2019}.

\section{Theoretical model}

\label{sec:model}

In this section, we introduce the emitter-waveguide model, the spatial
coupling profile, and the scattering-matrix formalism used to compute
the transport quantities.

\subsection{Emitter array and coupling profile}

\begin{figure}[h]
\begin{centering}
\includegraphics[width=8.8cm]{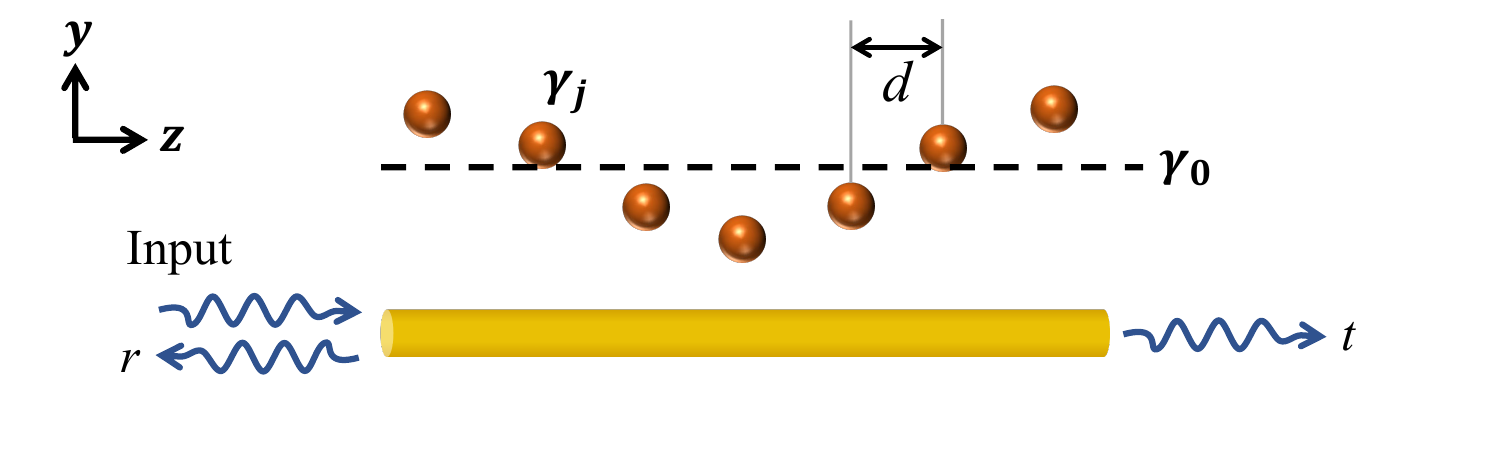} 
\par\end{centering}
\caption{Schematic illustration of a geometry-engineered atomic mirror in a
waveguide QED system. $N$ two-level emitters are coupled to a bidirectional
1D waveguide with uniform interatomic spacing $d$ along the propagation
axis ($z$ axis). The local spontaneous decay rates $\gamma_{j}$
are modulated by tailored transverse displacements of the emitters
relative to the waveguide axis. Because the guided mode is evanescent
in the transverse direction, such displacements modify the local coupling
strength exponentially. }
\label{fig:fig01} 
\end{figure}

We consider a waveguide QED system of $N$ two-level emitters coupled
to a bidirectional 1D waveguide, shown schematically in Fig.~\ref{fig:fig01}.
The emitters are placed along the propagation axis at positions $z_{j}=j\,d$,
where $j=1,2,\dots,N$ labels the emitters and $d$ is the uniform
interatomic spacing. Within the rotating-wave approximation, the Hamiltonian
$H=H_{\text{atom}}+H_{\text{field}}+H_{\text{int}}$ of the emitter-waveguide
system reads 
\begin{equation}
\begin{aligned}H= & \sum_{j=1}^{N}\omega_{0}b_{j}^{\dagger}b_{j}+\int dk\,\omega_{k}\left(c_{R}^{\dagger}(k)c_{R}(k)+c_{L}^{\dagger}(k)c_{L}(k)\right)\\
 & +\sum_{j=1}^{N}\int dk\left[g_{j}(k)b_{j}^{\dagger}c_{R}(k)e^{ikz_{j}}+g_{j}(k)b_{j}^{\dagger}c_{L}(k)e^{-ikz_{j}}+\text{H.c.}\right],
\end{aligned}
\label{eq:hamiltonian}
\end{equation}
where $b_{j}$ ($b_{j}^{\dagger}$) is the lowering (raising) operator
of the $j$-th emitter, whose bare transition frequency is $\omega_{0}$.
The operators $c_{R}(k)$ and $c_{L}(k)$ annihilate right- and left-propagating
waveguide photons of wavenumber $k.$ The photon dispersion is linearized
about the atomic resonance, $\omega_{k}\approx\omega_{0}+v_{g}(k-k_{0}),$
with $v_{g}$ the group velocity and $k_{0}=\omega_{0}/v_{g}$ the
resonant wavevector. Within the narrow frequency window of interest
we take the coupling to be frequency independent, $g_{j}\left(k\right)\approx g_{j}\left(k_{0}\right).$
We set $\hbar=1$ throughout and quote all frequencies in units of
the reference decay rate $\gamma_{0}$ defined below.

In the context of this model, we make several important assumptions.
First, we work within the rotating-wave approximation and linearize
the photon dispersion about the atomic resonance. Second, we assume
that the emitters radiate only into the guided mode, so that the array
is lossless and all transport is elastic. Any decay into non-guided
channels would enter as an additional loss channel and reduce the
peak reflectivity. Third, we neglect the propagation delay across
the array, $|z_{j}-z_{l}|/v_{g}\rightarrow0,$ which holds as long
as the retardation time $Nd/v_{g}$ is short compared with the collective
decay time. For macroscopically separated emitters this condition
fails and non-Markovian corrections modify the collective dynamics~\citep{Sinha2020,Zheng2013,Dinc2019}.
Finally, we restrict the analysis to the single-excitation sector,
appropriate to the weak monochromatic driving assumed throughout.

The geometry enters the model through the emitter-waveguide coupling.
Transverse displacement of an emitter relative to the waveguide axis
changes its local coupling exponentially, since the guided mode decays
evanescently away from the axis~\citep{LeKien2004,Goban2012}. The
same displacement has been used to control collective relaxation in
waveguide QED arrays~\citep{Lee2025}. We parameterize the transverse
position as 
\begin{equation}
y_{j}=y_{0}-2\eta\zeta\exp\left[-\frac{(\xi_{j}-1/2)^{2}}{2\sigma^{2}}\right],\label{eq:displacement_profile}
\end{equation}
where $y_{0}$ is a uniform offset, $\zeta$ is the evanescent decay
length, $\eta$ is a dimensionless tuning amplitude, $\sigma$ is
the dimensionless profile width, and $\xi_{j}=j/N$ is the normalized
coordinate along the array. Taking $g_{j}(k_{0})\propto\exp(-y_{j}/\zeta)$
and absorbing the uniform factor associated with $y_{0}$ into the
normalization gives the local decay rate 
\begin{equation}
\gamma_{j}=\frac{4\pi g_{j}\left(k_{0}\right)^{2}}{v_{g}}=\mathcal{N}_{D}\exp\left[4\eta\exp\left(-\frac{(\xi_{j}-1/2)^{2}}{2\sigma^{2}}\right)\right],\label{eq:gamma_profile}
\end{equation}
To ensure a fair comparison between different profiles, we fix the
global factor $\mathcal{N}_{D}$ by the normalization $\sum_{j=1}^{N}\gamma_{j}=N\gamma_{0}$,
where $\gamma_{0}$ is the decay rate of the corresponding uniform
array. This isolates the effect of the spatial distribution of the
coupling.

Applying the Heisenberg equations of motion to the field operators
and integrating them formally in time allows the photonic reservoir
to be eliminated~\citep{Gardiner1985,Lalumiere2013}. Within the
Markovian approximation set out above, this yields an effective non-Hermitian
Hamiltonian in the single-excitation subspace~\citep{AsenjoGarcia2017,Moiseyev2011,Ashida2020},
\begin{equation}
(H_{\text{eff}})_{jl}=\omega_{0}\delta_{jl}-\frac{i}{2}\sqrt{\gamma_{j}\gamma_{l}}\,e^{i\phi_{0}|j-l|},\label{eq:Heff}
\end{equation}
where $\phi_{0}=k_{0}d$ is the interatomic phase, i.e., the propagation
phase accumulated over one lattice spacing at the atomic resonance.
The Bragg and anti-Bragg conditions correspond to $\phi_{0}=0,\pi$
and $\phi_{0}=\pi/2$, respectively.

\subsection{Scattering matrix formalism}

We compute the transmission and reflection amplitudes $t$ and $r$
from the $S$-matrix~\citep{Gardiner1985,Caneva2015,Xu2015,Shi2015}.
Under weak monochromatic driving the emitter amplitudes oscillate
at the incident frequency $\omega_{\text{in}},$ with $b_{j}(t)=b_{j}e^{-i\omega_{\text{in}}t}$,
and the steady state is $\mathbf{b}=G\mathbf{u}_{R},$ where $\mathbf{b}$
collects the $b_{j}$ normalized to the incident field amplitude and
the inverse Green's matrix $G^{-1}\equiv\omega_{\text{in}}I-H_{\text{eff}},$
with elements 
\begin{equation}
(G^{-1})_{jl}=\Delta\delta_{jl}+\frac{i}{2}\sqrt{\gamma_{j}\gamma_{l}}\,e^{i\phi_{0}|j-l|},\label{eq:Ginv}
\end{equation}
with $\Delta\equiv\omega_{\text{in}}-\omega_{0}$ the probe detuning
used throughout. The drive enters through $(\mathbf{u}_{R})_{j}=\frac{1}{\sqrt{2}}\sqrt{\gamma_{j}}e^{i\phi_{0}j},$
and $(\mathbf{u}_{L})_{j}=\frac{1}{\sqrt{2}}\sqrt{\gamma_{j}}e^{-i\phi_{0}j}$
projects the emitter amplitudes onto the backward channel. The transmitted
field is the incident field plus what the array radiates forward,
while the reflected field is radiated backward only, 
\begin{equation}
t=1-i\,\mathbf{u}_{R}^{\dagger}G\mathbf{u}_{R},\qquad r=-i\,\mathbf{u}_{L}^{\dagger}G\mathbf{u}_{R},\label{eq:tr}
\end{equation}
and $|t|^{2}+|r|^{2}=1$ at arbitrary detuning, since the guided mode
is the only available channel. Equation~(\ref{eq:tr}) is exact within
the Markovian single-excitation model.

\section{Reflection spectra and bandwidth}

\label{sec:numerics}

In this section, we investigate how the spatial coupling profile modifies
the reflection spectrum and the high-reflectivity bandwidth.

\subsection{Reflection spectra}

We first characterize the transport response by computing the single-photon
reflection spectrum $|r|^{2}$ as a function of the detuning $\Delta.$
We take a representative finite array of $N=20$ emitters and fix
the interatomic phase at the anti-Bragg condition $\phi_{0}=0.5\pi.$
To isolate the response of the reflection window to the coupling profile,
we hold the width at $\sigma=0.1$ and evaluate the spectrum for several
tuning amplitudes $\eta$. 
\begin{figure}[h]
\centering %
\begin{tabular}{@{}l@{}}
(a)\tabularnewline
\includegraphics[width=6cm]{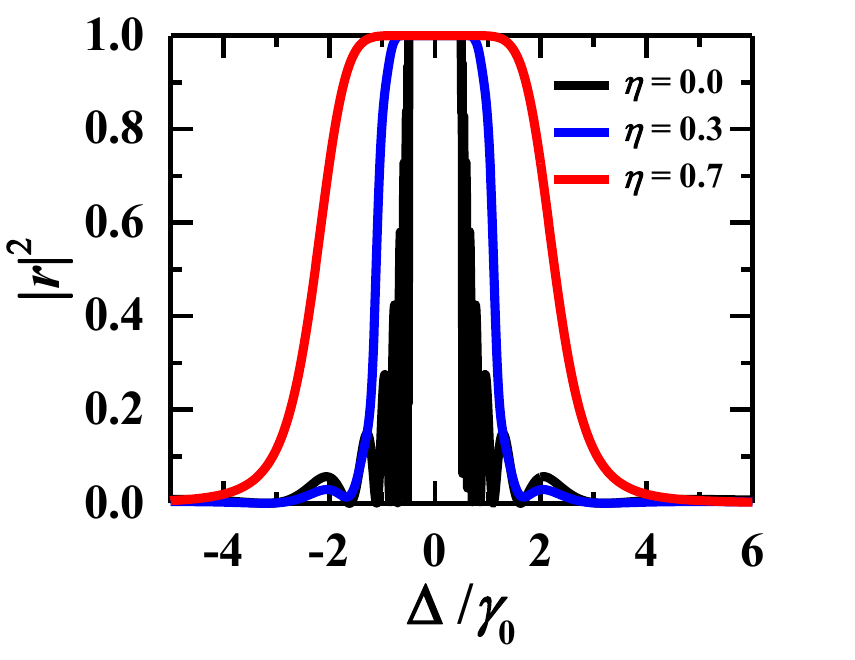}\tabularnewline
(b)\tabularnewline
\includegraphics[width=6cm]{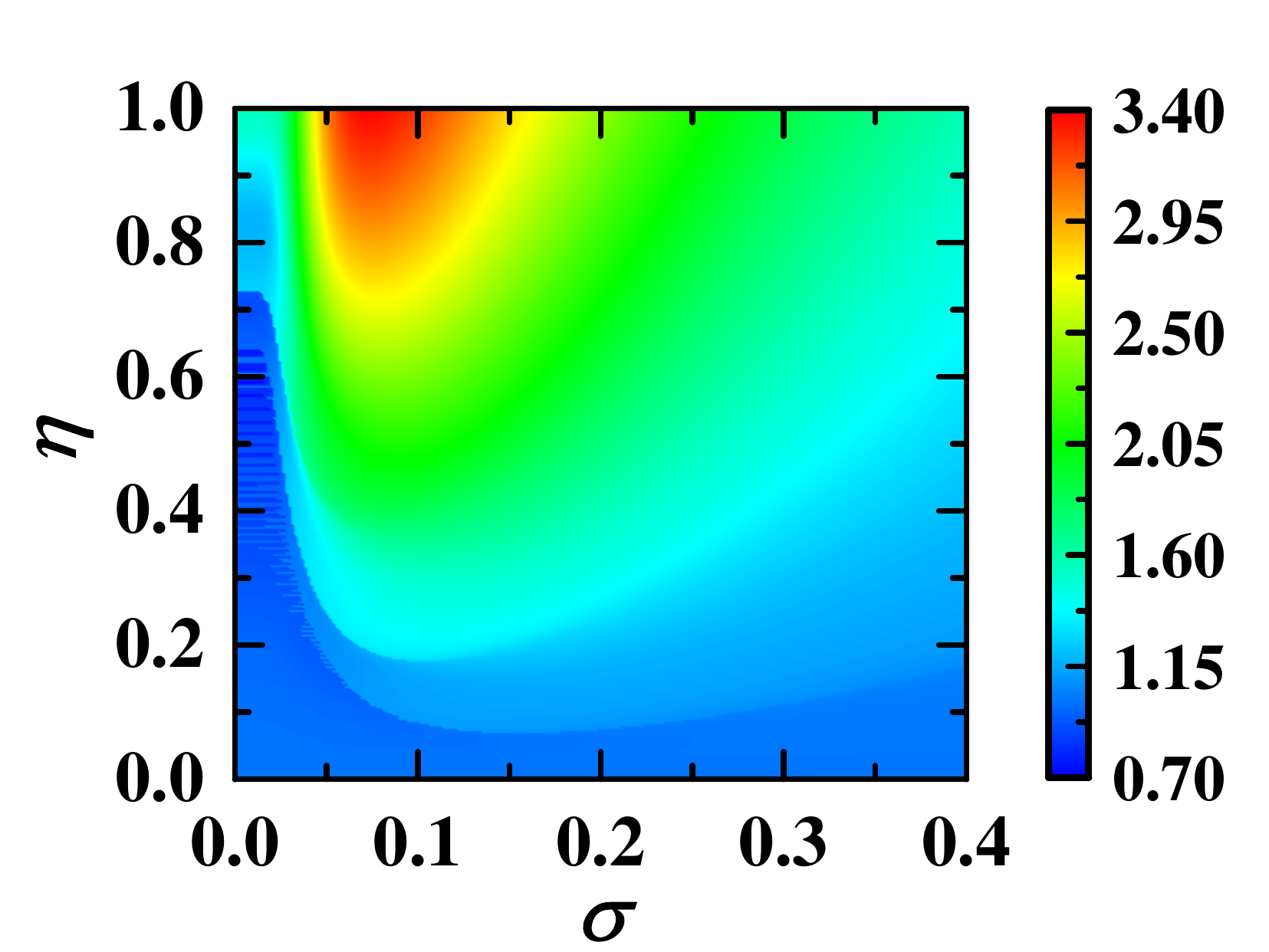}\tabularnewline
\end{tabular}\caption{(a) Single-photon reflection spectrum $|r|^{2}$ as a function of
the detuning $\Delta$ for different tuning amplitudes $\eta$ at
$\sigma=0.1$. The black curve is the uniform array ($\eta=0$). The
blue and red curves correspond to coupling-engineered configurations
with $\eta=0.3$ and $\eta=0.7$, respectively. (b) High-reflectivity
bandwidth (defined by $|r|^{2}>0.99$) mapped across the two-dimensional
parameter space of the tuning amplitude $\eta$ and the profile width
$\sigma$. The color scale gives the continuous width of the flat
near-unity reflection window in units of $\gamma_{0}$. In both panels,
$N=20$ and $\phi_{0}=0.5\pi$. }
\label{fig:fig02} 
\end{figure}

Figure~\ref{fig:fig02}(a) shows the main effect of the spatial modulation.
For the uniform array ($\eta=0$), the reflection spectrum is confined
by the bare collective bandgap and carries pronounced Fabry-Perot
oscillations at the edges of the reflection band, which we attribute
to the abrupt impedance step at the two ends of the array. As $\eta$
increases from $0.3$ to $0.7$, the peak reflection region broadens,
and these oscillations are progressively smoothed into a flat plateau
with $|r|^{2}>0.99.$ We take up the mechanism behind both in Sec.~\ref{sec:spectral}
and quantify the widening in Sec.~\ref{sec:scaling}.

Quantitatively, the $S$-matrix calculation gives a $|r|^{2}>0.99$
window of $1.00\,\gamma_{0}$ for the uniform array, which widens
to $1.51\,\gamma_{0}$ at $\eta=0.3$ and $2.62\,\gamma_{0}$ at $\eta=0.7$.
This corresponds to an enhancement by a factor of $2.6$.

\subsection{Bandwidth over the envelope parameters}

To examine the mirror performance over the full envelope parameter
space, we compute the high-reflectivity bandwidth---defined as the
continuous spectral window over which $|r|^{2}>0.99$---across the
plane spanned by the tuning amplitude $\eta$ and the profile width
$\sigma$.

Along the line $\eta\rightarrow0$ in Fig.~\ref{fig:fig02}(b), the
system reduces to the uniform limit and the bandwidth takes its bare
value independently of $\sigma.$ Increasing $\eta$ extends the reflection
window, but the broadening is bounded on both sides by the width $\sigma.$
A very narrow envelope spans less than one lattice spacing, so the
normalization concentrates a large fraction of the coupling on one
or a few central emitters and weakens the coupling of the remaining
sites. In the opposite limit, a very broad envelope spreads the coupling
almost evenly across the array, and the modulation has little effect.
As a result, the widest reflection window appears along a ridge at
intermediate $\sigma$. We return to the origin of this ridge in Sec.~\ref{sec:scaling}.

\section{Spectral factorization and pole reorganization}

\label{sec:spectral}

In this section, we identify the microscopic origin of the broadening
and of the suppressed Fabry-Perot oscillations found in Sec.~\ref{sec:numerics}.
We analyze the zeros and poles of the scattering matrix. For the sharply
localized profile of Fig.~\ref{fig:fig02}, only a few of the $N=20$
emitters lie within the strongly coupled region. A slowly varying
local-periodic description is therefore not controlled. We instead
use a spectral factorization of the scattering matrix, based on the
determinant ratios derived in Appendix~\ref{app:det}. The factorization
is exact at any $N$ and never invokes a band edge. Zeros and poles
in the complex frequency plane have been used to characterize reflectionless
and perfectly absorbing configurations in open photonic systems~\citep{Moiseyev2011,Sweeney2020}.
The propagation phase in $H_{\text{eff}}$ is fixed at $\phi_{0}$,
so both determinants below are polynomials of degree $N$ in $\omega_{\text{in}}$
and factor over their complex roots, 
\begin{align}
\det\left(A_{t}(\omega_{\text{in}})\right) & =\prod_{n=1}^{N}(\omega_{\text{in}}-\zeta_{n}),\nonumber \\
\det\left(G^{-1}(\omega_{\text{in}})\right) & =\prod_{n=1}^{N}(\omega_{\text{in}}-p_{n}),\label{eq:factorization}
\end{align}
where $A_{t}\equiv G^{-1}-i\mathbf{u}_{R}\mathbf{u}_{R}^{\dagger}$
is the modified forward-scattering matrix introduced in Appendix~\ref{app:det},
the $\zeta_{n}$ are the algebraic zeros of its determinant and become
physical transmission zeros when they are not canceled by the $p_{n}$,
and the $p_{n}$ are the collective poles, i.e., the complex eigenvalues
of $H_{\text{eff}},$ which describe the dissipative resonances of
the open array, with $n=1,2\dots,N$ labelling the collective modes.
Substituting Eq.~(\ref{eq:factorization}) into the transport relations
gives a spectral reconstruction in which the total transmission is
a product of $N$ single-pole factors, 
\begin{equation}
T(\omega_{\text{in}})=|t|^{2}=\prod_{n=1}^{N}\frac{|\omega_{\text{in}}-\zeta_{n}|^{2}}{|\omega_{\text{in}}-p_{n}|^{2}}\equiv\prod_{n=1}^{N}F_{n}.\label{eq:product}
\end{equation}
A single factor $F_{n}\rightarrow0$ suppresses $T$ entirely, whereas
a mode with $F_{n}\approx1$ leaves the spectrum unchanged. A wide
window of high reflectivity therefore requires small factors and no
strongly amplifying factor that remains uncompensated by the others.
The analysis below focuses on how the pole reorganization suppresses
such amplification.

\subsection{Degeneracy of the zeros and Bragg cancellation}

The numerator of Eq.~(\ref{eq:product}) can be evaluated in closed
form. As shown in Appendix~\ref{app:det}, the rank-one term $-i\mathbf{u}_{R}\mathbf{u}_{R}^{\dagger}$
cancels the Markovian kernel $e^{i\phi_{0}|j-l|}$ exactly on the
triangle $j\geq l$, where $|j-l|=j-l$. The matrix $A_{t}$ carries
the constant diagonal $\Delta$, and $A_{t}-\Delta I$ is strictly
upper triangular and hence nilpotent, giving 
\begin{equation}
\det\left(A_{t}(\omega_{\text{in}})\right)=\Delta^{N},\label{eq:zeros_exact}
\end{equation}
for any positive coupling profile $\{\gamma_{j}\}$, any $N$, and
any $\phi_{0}$ within the present reciprocal, lossless Markovian
model. Equation~(\ref{eq:zeros_exact}) places all $N$ algebraic
zeros of the numerator at the atomic resonance, independently of the
coupling modulation. For strictly positive couplings and a non-Bragg
phase $0<\phi_{0}<\pi$, $\det[G^{-1}(\omega_{0})]\neq0$, so these
roots are not canceled by collective poles and are physical transmission
zeros. In this regime, substitution into Eq.~(\ref{eq:product})
gives the exact form 
\begin{equation}
T(\omega_{\text{in}})=\frac{(\omega_{\text{in}}-\omega_{0})^{2N}}{\prod_{n=1}^{N}|\omega_{\text{in}}-p_{n}|^{2}},\qquad F_{n}=\frac{\Delta^{2}}{|\Delta-\tilde{p}_{n}|^{2}},\label{eq:T_exact}
\end{equation}
with $\tilde{p}_{n}=p_{n}-\omega_{0}$ the pole positions measured
from the atomic resonance.

The exact Bragg phases $\phi_{0}=0,\pi$ form an important exception.
There, $H_{\mathrm{eff}}-\omega_{0}I$ has rank one, with one bright
pole at $\omega_{0}-i\sum_{j}\gamma_{j}/2=\omega_{0}-iN\gamma_{0}/2$
and $N-1$ dark poles at $\omega_{0}$. Consequently, 
\begin{equation}
\det(G^{-1})=\Delta^{N-1}\left(\Delta+iN\gamma_{0}/2\right),\qquad t=\frac{\Delta}{\Delta+iN\gamma_{0}/2},\label{eq:bragg_cancellation}
\end{equation}
so $N-1$ common factors cancel and the physical transmission amplitude
has only a simple zero. Away from the Bragg points, the uncanceled
$N$-fold amplitude zero guarantees perfect on-resonance reflection
and a flat-bottomed transmission dip whose flatness increases with
$N$. Within the present lossless Markovian model, the algebraic numerator
is profile independent, while coupling engineering changes the bandwidth
through the collective poles.

\subsection{Pole reorganization for $N=20$}

\begin{figure}[h]
\begin{centering}
\includegraphics[width=8.8cm]{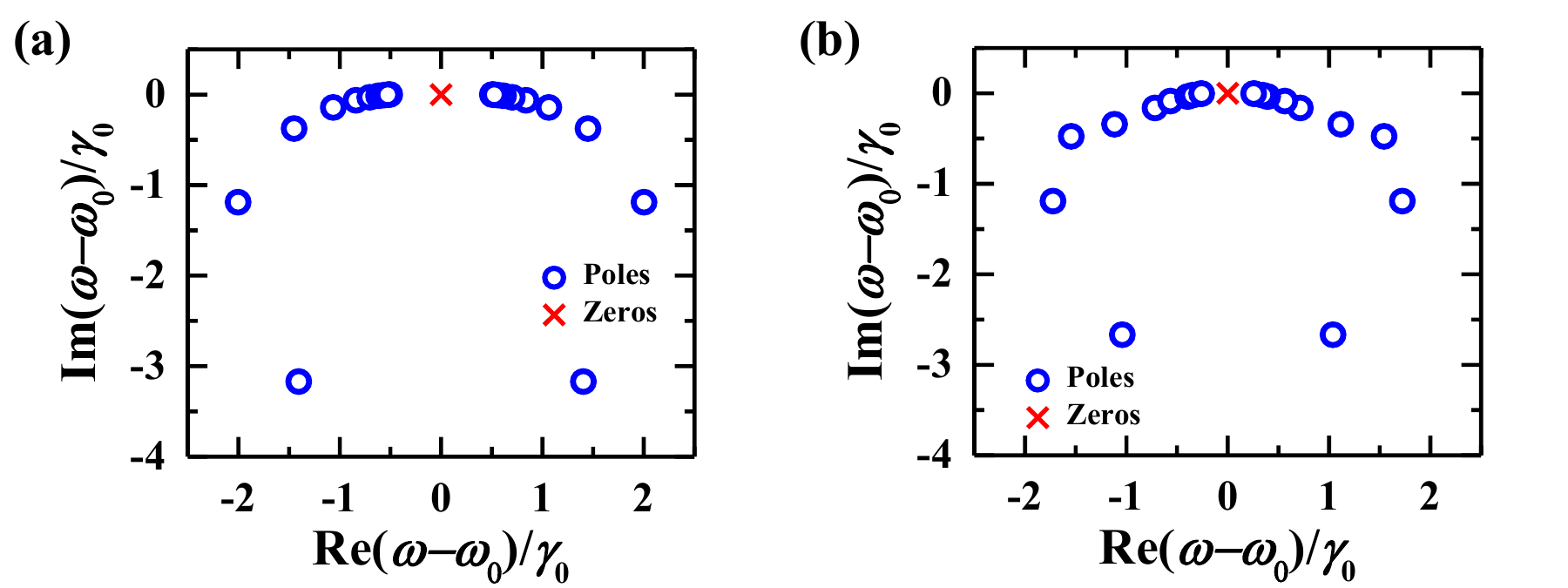} 
\par\end{centering}
\caption{Distribution of the collective poles $\ensuremath{\tilde{p}_{n}=p_{n}-\omega_{0}}$
(blue circles) in the complex frequency plane for $N=20$ and $\phi_{0}=0.5\pi$.
The $N$ transmission zeros are exactly degenerate at the origin {[}Eq.~(\ref{eq:zeros_exact}){]}
and are indicated by a single marker. (a) Uniform array ($\eta=0$):
the deepest superradiant pole lies at $\text{Im}(\tilde{p})/\gamma_{0}\approx-3.17$,
while the subradiant poles crowd against the real axis within the
narrow window $0.51\lesssim|\text{Re}(\tilde{p})|/\gamma_{0}\lesssim0.56$
and pin the window edge there. (b) Coupling-engineered array ($\eta=0.5$,
$\sigma=0.1$): the radiative width is shared more evenly, the deepest
pole is raised to $\text{Im}(\tilde{p})/\gamma_{0}\approx-2.67$ and
the shallowest is lifted by an order of magnitude, and no single mode
limits the band. }
\label{fig:zeros_20} 
\end{figure}

We now examine the distribution of the collective poles for $N=20$
and relate the broadening of the reflection window in Fig.~\ref{fig:fig02}(a)
to the pole configuration. In the uniform array {[}Fig.~\ref{fig:zeros_20}(a){]},
the radiative width is distributed unevenly: the most dissipative
superradiant pole reaches $\text{Im}(\tilde{p})/\gamma_{0}\approx-3.17$,
while six subradiant poles crowd against the real axis with $|\text{Im}(\tilde{p})|/\gamma_{0}<10^{-2}$.
Their real parts also cluster within the narrow interval $0.51\lesssim|\text{Re}(\tilde{p})|/\gamma_{0}\lesssim0.56$.

This crowding limits the bandwidth of the uniform mirror. Writing
$\tilde{p}_{n}=x_{n}-iy_{n}$ where $y_{n}>0$ is the radiative width
of the $n$th collective mode, Eq.~(\ref{eq:T_exact}) shows that
the factor associated with each pole grows to $x_{n}^{2}/y_{n}^{2}$
as the probe detuning sweeps through $\Delta=x_{n}$. For a mode with
$y_{n}=6\times10^{-4}\gamma_{0}$, this factor becomes extremely large,
so the transmission reaches the threshold $T=0.01$ (i.e., $|r|^{2}=0.99$)
as soon as the probe arrives at the innermost subradiant resonance,
i.e., the one with the smallest $|\text{Re}(\tilde{p})|$. The window
edge is pinned to this resonance, and the bandwidth is given by twice
its detuning from the atomic resonance. We find that the two agree
to within $1\%$ for all system sizes examined between $N=20$ and
$N=80$. At $N=20$, for example, the exact window is $1.00\gamma_{0}$,
compared with $2|\text{Re}(\tilde{p})|_{\text{min}}=1.01\gamma_{0}$.
The reflection window of the uniform array is terminated abruptly
by a single subradiant mode.

When the Gaussian profile is applied {[}$\eta=0.5$, $\sigma=0.1$,
Fig.~\ref{fig:zeros_20}(b){]}, this pinning is lifted. Although
the innermost pole moves inward, from $|\text{Re}(\tilde{p})|/\gamma_{0}=0.51$
to $0.26$, the reflection window now extends to $\Delta=1.05\gamma_{0}$,
about four times farther than this pole. At $N=80$, this ratio increases
to $6.4$.

This enhancement originates from the equalization of the radiative
widths rather than from a displacement of the resonances. The shallowest
pole is lifted from $\text{Im}(\tilde{p})/\gamma_{0}=-6.3\times10^{-4}$
to $-4.3\times10^{-3}$, and the deepest one is raised from $-3.17$
to $-2.67$, so the spread of radiative widths narrows from a factor
of $5\times10^{3}$ to $6\times10^{2}$. Accordingly, the largest
ratio of detuning to radiative width, which sets the peak value of
each factor through Eq.~(\ref{eq:T_exact}), decreases from $805$
to $60$. As a result, no single mode dominates the window edge, which
is instead determined collectively by several modes with moderate
factors. The same equalization of the radiative widths provides a
consistent explanation for the suppression of the boundary-induced
Fabry-Perot oscillations in Fig.~\ref{fig:fig02}(a).

Note that the widening does not result from pushing the poles away
from the atomic resonance, since the innermost pole in fact moves
inward. This indicates that the bandwidth is governed by how evenly
the radiative width is shared among the collective modes rather than
by the positions of the resonances. A uniform array concentrates most
of the radiative width in a few superradiant modes and leaves the
remaining modes nearly dark, and these dark modes terminate the reflection
window.

\subsection{Single-mode factors in a four-emitter array}

\begin{figure}[h]
\begin{centering}
\includegraphics[width=8.8cm]{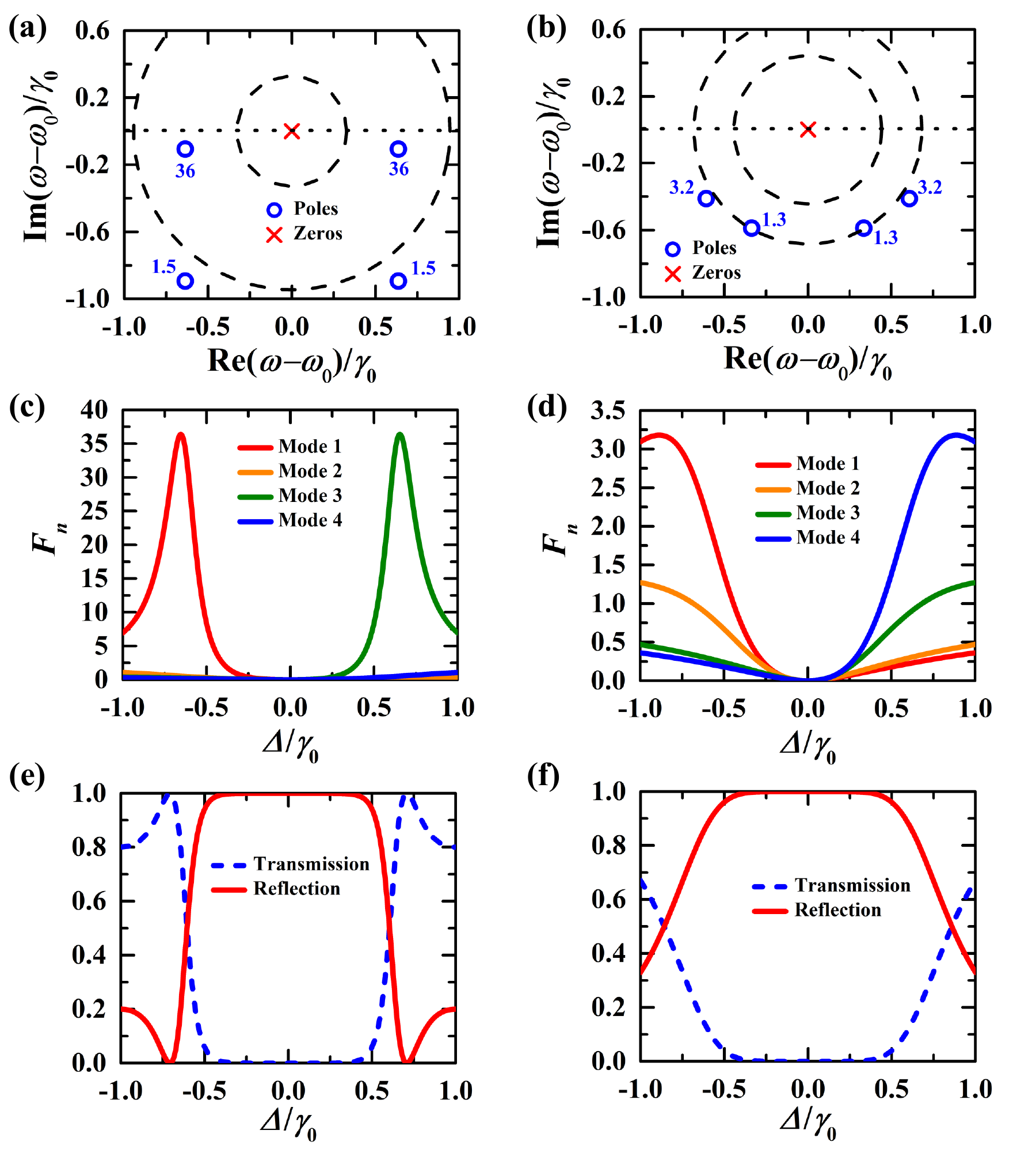} 
\par\end{centering}
\caption{Pole distribution and spectral reconstruction for a toy-model system
($N=4$) at the anti-Bragg condition ($\phi_{0}=0.5\pi$). (a),(b)
Complex-plane distributions of the collective poles $\ensuremath{\tilde{p}_{n}=p_{n}-\omega_{0}}$
(blue circles) for (a) the uniform array and (b) the coupling-engineered
Gaussian array ($\eta=0.5$, $\sigma=0.25$). The $N$-fold degenerate
transmission zero at the origin is marked by a single red cross. Dashed
circles indicate the non-amplifying criterion of Eq.~(\ref{eq:noamp}).
(c),(d) Individual mode factors $F_{n}(\Delta)$ for (c) the uniform
array, in which the two near-real subradiant poles leave the plotted
range at $|\Delta|/\gamma_{0}\approx0.48$ and peak at $F_{n}^{\text{max}}=36$,
and (d) the Gaussian array, in which no factor exceeds $3.2$ and
none exceeds $1.4$ within $|\Delta|<0.5\gamma_{0}$. (e),(f) Reconstructed
total transmission $T$ (blue dashed) and reflection $|r|^{2}$ (red
solid) obtained from the product of Eq.~(\ref{eq:product}) for (e)
the uniform array and (f) the coupling-engineered profile. }
\label{fig:zeros_4} 
\end{figure}

To illustrate how this pole configuration shapes the reflection spectrum,
we consider a toy model with $N=4$ and a relaxed profile width $\sigma=0.25$,
chosen so that neighboring resonances do not overlap and each factor
$F_{n}$ can be tracked separately. In terms of $x_{n}$ and $y_{n}$,
Eq.~(\ref{eq:T_exact}) gives 
\begin{equation}
F_{n}(\Delta)=\frac{\Delta^{2}}{(\Delta-x_{n})^{2}+y_{n}^{2}},\label{eq:Fn}
\end{equation}
so that $F_{n}<1$---the mode is non-amplifying---provided 
\begin{equation}
|\Delta|<\Delta_{n}^{*}\equiv\frac{x_{n}^{2}+y_{n}^{2}}{2|x_{n}|}=\frac{|\tilde{p}_{n}|^{2}}{2|\text{Re}\,\tilde{p}_{n}|}.\label{eq:noamp}
\end{equation}
Equation~(\ref{eq:Fn}) also gives the maximum of each factor: $F_{n}$
attains its largest value at $\Delta=2\Delta_{n}^{*}$, where 
\begin{equation}
F_{n}^{\text{max}}=\frac{|\tilde{p}_{n}|^{2}}{y_{n}^{2}}=1+\left(\frac{\text{Re}\,\tilde{p}_{n}}{\text{Im}\,\tilde{p}_{n}}\right)^{2}.\label{eq:Fmax}
\end{equation}
Equations~(\ref{eq:noamp}) and (\ref{eq:Fmax}) show that the edges
of the reflection window are controlled by a single dimensionless
quantity, namely the ratio of a resonance's detuning from the atomic
resonance to its radiative width. A mode does not amplify the transmission
as long as the probe detuning stays within $\pm\Delta_{n}^{*},$ indicated
in Fig.~\ref{fig:zeros_4}(a), (b) by a dashed circle of radius $\Delta_{n}^{*}$
about the origin. A larger radiative width enlarges this circle. For
a small array, in which there are too few strongly suppressed factors
to compensate, $\min_{n}\Delta_{n}^{*}$ sets the detuning at which
the first factor begins to amplify.

In the uniform limit ($\eta=0$), the pole configuration of Fig.~\ref{fig:zeros_4}(a)
places the collective poles at $\tilde{p}_{n}/\gamma_{0}=\pm0.64-0.11i$
and $\pm0.64-0.89i$. The long-lived pair lies close to the real axis,
and Eq.~(\ref{eq:noamp}) gives $\Delta_{n}^{*}=0.33\gamma_{0}$
for this pair. As the probe detuning exceeds this value, the real-part
separation $(\Delta-x_{n})^{2}$ decreases, and the denominator of
Eq.~(\ref{eq:Fn}) is limited only by the small imaginary part $y_{n}=0.11\gamma_{0}$.
The corresponding factor rises steeply. It crosses unity at $|\Delta|=0.33\gamma_{0}$,
leaves the plotted range at $|\Delta|\approx0.48\gamma_{0}$ {[}Fig.~\ref{fig:zeros_4}(c){]},
and reaches its maximum $F_{n}^{\text{max}}=36$ at $|\Delta|=0.65\gamma_{0}$
according to Eq.~(\ref{eq:Fmax}). Through the product $T=\prod_{n}F_{n}$,
this single enhanced factor degrades the reflection spectrum in Fig.~\ref{fig:zeros_4}(e)
and rounds off the reflection window prematurely.

The Gaussian profile ($\eta=0.5$) suppresses this enhancement by
modifying both terms in the denominator of Eq.~(\ref{eq:Fn}) simultaneously.
As Fig.~\ref{fig:zeros_4}(b) shows, the modulation redistributes
the four resonances to $\tilde{p}_{n}/\gamma_{0}=\pm0.61-0.41i$ and
$\pm0.34-0.59i$. The near-real subradiant pair of the uniform array
no longer exists, and every mode now has a radiative width comparable
to the bandwidth. The corresponding thresholds are $\Delta_{n}^{*}=0.44\gamma_{0}$
and $0.68\gamma_{0}$, so the narrowest non-amplifying range widens
from $0.65\gamma_{0}$ to $0.89\gamma_{0}$, corresponding to an enhancement
of $36\%$. By Eq.~(\ref{eq:Fmax}), the worst-case factor drops
from $F_{n}^{\text{max}}=36$ to $3.2$.

As a result, $|\Delta-\tilde{p}_{n}|^{2}$ remains bounded from below
for all real probe detunings. No factor exceeds $3.2$ across the
plotted range, and within $|\Delta|<0.5\gamma_{0}$ all four factors
remain below $1.4$ {[}Fig.~\ref{fig:zeros_4}(d){]}. Their product
yields the reflection window in Fig.~\ref{fig:zeros_4}(f), whose
edges decrease smoothly instead of showing the collapse and rebound
in Fig.~\ref{fig:zeros_4}(e). Since the depth of the transmission
minimum is fixed by the $N$-fold amplitude zero, or equivalently
the $2N$-th order zero of $T$, at resonance {[}Eq.~(\ref{eq:zeros_exact}){]},
it cannot be optimized. Coupling engineering controls only the flatness
of the window edges, through the radiative widths and the real-axis
spacing of the collective poles.

\section{Semiclassical finite-size scaling}

\label{sec:scaling}

We now examine the finite-size dependence of the high-reflectivity
bandwidth $w(N)$. The spectral factorization of Sec.~\ref{sec:spectral}
explains the enhancement at a given array size, but it yields the
pole positions only numerically, one configuration at a time. To obtain
the explicit dependence of $w(N)$ on the emitter number $N$, on
the envelope parameters, and on the interatomic phase $\phi_{0}$.
In a finite array, the departure from the infinite-lattice limit arises
from the residual impedance mismatch at the boundaries and the evanescent
leakage through the gap. A semiclassical tunneling treatment based
on the WKB approximation~\citep{Bender1999} yields an algebraic
scaling law for the convergence of the reflection window toward its
bulk asymptote.

\subsection{Semiclassical scaling function}

We map single-photon transport near the edges of the collective bandgap
onto tunneling through an effective potential barrier formed by the
spatially modulated decay rate. Provided the envelope varies slowly
over one lattice spacing, $\sigma N\gg1$, the array is locally uniform
and a WKB treatment applies. We replace the discrete label $\xi_{j}$
of Eq.~(\ref{eq:gamma_profile}) by a continuous coordinate and write
the envelope as $\gamma\left(\xi\right),$ or as $\gamma\left(z\right)$
with $z=\xi L$ and $L=Nd$ the array length. Evaluating the attenuation
phase $\theta$ accumulated across the barrier and imposing the threshold
$|r|^{2}=0.99$ through the Kemble transmission formula~\citep{Kemble1935}
reduces the bandwidth to the compact form 
\begin{equation}
w(N)=w_{\infty}+\frac{\Lambda}{N},\label{eq:scaling}
\end{equation}
with $w_{\infty}$ the bulk bandwidth, i.e., the limit of $w(N)$
for $N\to\infty,$ and $\Lambda$ the finite-size amplitude. The derivation,
including the local dispersion relation, the Taylor expansion of the
barrier, and the unit-cell transfer matrix $\tilde{M}$, is given
in Appendix~\ref{app:wkb}. Solving the band-edge condition $\text{Tr}(\tilde{M})=\pm2$
gives the bulk bandwidth, 
\begin{equation}
w_{\infty}(\eta,\sigma,\phi_{0})=\frac{\gamma_{\text{max}}(\eta,\sigma)}{\sin\phi_{0}}.\label{eq:winf}
\end{equation}
Here $\gamma_{\text{max}}=\max_{j}\gamma_{j}$ is the peak decay rate
at the array center. The bulk bandwidth is set by the largest local
rate and is independent of the array average. Eliminating $\mathcal{N}_{D}$
through the normalization $\sum_{j}\gamma_{j}=N\gamma_{0}$ and passing
to the continuum, $\sum_{j}\to N\int_{0}^{1}d\xi,$ gives it in closed
form, 
\begin{equation}
\frac{\gamma_{\text{max}}}{\gamma_{0}}=\frac{e^{4\eta}}{\int_{0}^{1}d\xi\,\exp\left[4\eta\,e^{-(\xi-1/2)^{2}/2\sigma^{2}}\right]},\label{eq:gmax}
\end{equation}
bounded between unity for $\sigma\gg1$ and $e^{4\eta}$ for $\sigma\to0.$
A peaked envelope yields $\gamma_{\text{max}}>\gamma_{0},$ and the
bandgap of the engineered array exceeds that of the uniform array.

The finite-size amplitude follows from the attenuation phase accumulated
near the two edges of the collective bandgap and is governed by the
curvature of the decay-rate profile at the array center. Expanding
Eq.~(\ref{eq:gamma_profile}) about $\xi=1/2$ gives 
\begin{equation}
\gamma(\xi)\simeq\gamma_{\text{max}}\exp\left[-\frac{(\xi-1/2)^{2}}{2\sigma_{\text{eff}}^{2}}\right],\qquad\sigma_{\text{eff}}=\frac{\sigma}{2\sqrt{\eta}}.\label{eq:sigma_eff}
\end{equation}
The barrier is controlled by $\sigma_{\text{eff}}$ rather than $\sigma$,
and a larger tuning amplitude sharpens the profile even at fixed $\sigma.$
In terms of $\sigma_{\text{eff}},$ 
\begin{equation}
\Lambda=\frac{\gamma_{\text{max}}\ln(1-|r|_{\text{th}}^{2})}{4\sqrt{2}\,\pi\sigma_{\text{eff}}}\left[\frac{\cos(\phi_{0}/2)}{\sin^{2}(\phi_{0}/2)}+\frac{\sin(\phi_{0}/2)}{\cos^{2}(\phi_{0}/2)}\right],\label{eq:lambda}
\end{equation}
where $|r|_{\text{th}}^{2}=0.99$ is the threshold that defines the
edge of the reflection window. Since $\ln(1-|r|_{\text{th}}^{2})<0$,
$\Lambda$ is negative and the bandwidth contracts at finite $N.$

The magnitude of $\Lambda$ depends on the envelope through $|\Lambda|\propto\gamma_{\text{max}}\sqrt{\eta}/\sigma,$
so a narrow or strongly modulated envelope leads to a larger finite-size
amplitude. The factor $\sqrt{\eta}$ shows that the coefficient of
this parabolic-barrier $1/N$ correction tends to zero as $\eta\to0$.
This limit does not imply the absence of a bandgap or of finite-size
effects in a uniform array: $\sigma_{\text{eff}}$ diverges, the turning-point
construction ceases to apply, and the residual uniform-array correction
lies beyond the present leading-order result. For $\eta>0$, combining
this result with Eq.~(\ref{eq:winf}) provides a qualitative picture
of the ridge in Fig.~\ref{fig:fig02}(b): the relative finite-size
correction 
\begin{equation}
\frac{|\Lambda|}{Nw_{\infty}}=\frac{\sqrt{2\eta}\,\left|\ln(1-|r|_{\text{th}}^{2})\right|\sin\phi_{0}}{4\pi\sigma N}\left[\frac{\cos(\phi_{0}/2)}{\sin^{2}(\phi_{0}/2)}+\frac{\sin(\phi_{0}/2)}{\cos^{2}(\phi_{0}/2)}\right]\label{eq:penalty}
\end{equation}
grows as $\sqrt{\eta}/(\sigma N)$, whereas the bulk bandwidth $w_{\infty}\propto\gamma_{\text{max}}$
saturates at $e^{4\eta}\gamma_{0}$ by Eq.~(\ref{eq:gmax}). At fixed
$\eta,$ this competition selects an optimal width $\sigma$, which
is expected to move to smaller values as $N$ increases. In contrast,
no such optimum exists for $\eta,$ so the competition determines
only the optimal $\sigma$. Note that Fig.~\ref{fig:fig02}(b) ($N=20$)
lies outside the regime of validity discussed below, so this picture
is only qualitative there.

Equation~(\ref{eq:penalty}) also determines the validity range of
the expansion. The product of $\sin\phi_{0}$ with the bracket is
invariant under $\phi_{0}\to\pi-\phi_{0}$, the two terms in the bracket
exchanging roles, so this correction is symmetric about the anti-Bragg
point and diverges at both edges, giving 
\begin{equation}
\frac{|\Lambda|}{Nw_{\infty}}\;\xrightarrow[\phi_{0}\to0,\pi]{}\;\frac{\sqrt{2\eta}\,\left|\ln(1-|r|_{\text{th}}^{2})\right|}{\pi\,\sigma N\min(\phi_{0},\pi-\phi_{0})}\;\simeq\;\frac{1.47}{\sigma N\min(\phi_{0},\pi-\phi_{0})}\label{eq:validity}
\end{equation}
for $\eta=0.5$. The finite-size expansion is controlled by $1/[\sigma N\min(\phi_{0},\pi-\phi_{0})]$
rather than by $1/N,$ and the scaling law requires 
\begin{equation}
\sigma N\min(\phi_{0},\pi-\phi_{0})\gg1,\label{eq:criterion}
\end{equation}
a condition that tightens as either Bragg point is approached. Two
further approximations fail in the same limit, although only after
Eq.~(\ref{eq:criterion}) has already been violated: (i) the bulk
expression $w_{\infty}=\gamma_{\text{max}}/\sin\phi_{0}$ diverges
as $\phi_{0}\to0,\pi$, whereas the true bandwidth cannot exceed the
total radiative width $N\gamma_{0}$ that the array draws from the
guided mode; and (ii) the turning points of the semiclassical barrier
move outward until the parabolic approximation, Eq.~(\ref{eq:parabolic})
in Appendix~\ref{app:wkb}, no longer holds. Equation~(\ref{eq:criterion})
provides the relevant validity condition.

\subsection{Comparison with exact results}

\begin{figure}[h]
\begin{centering}
\includegraphics[width=8.8cm]{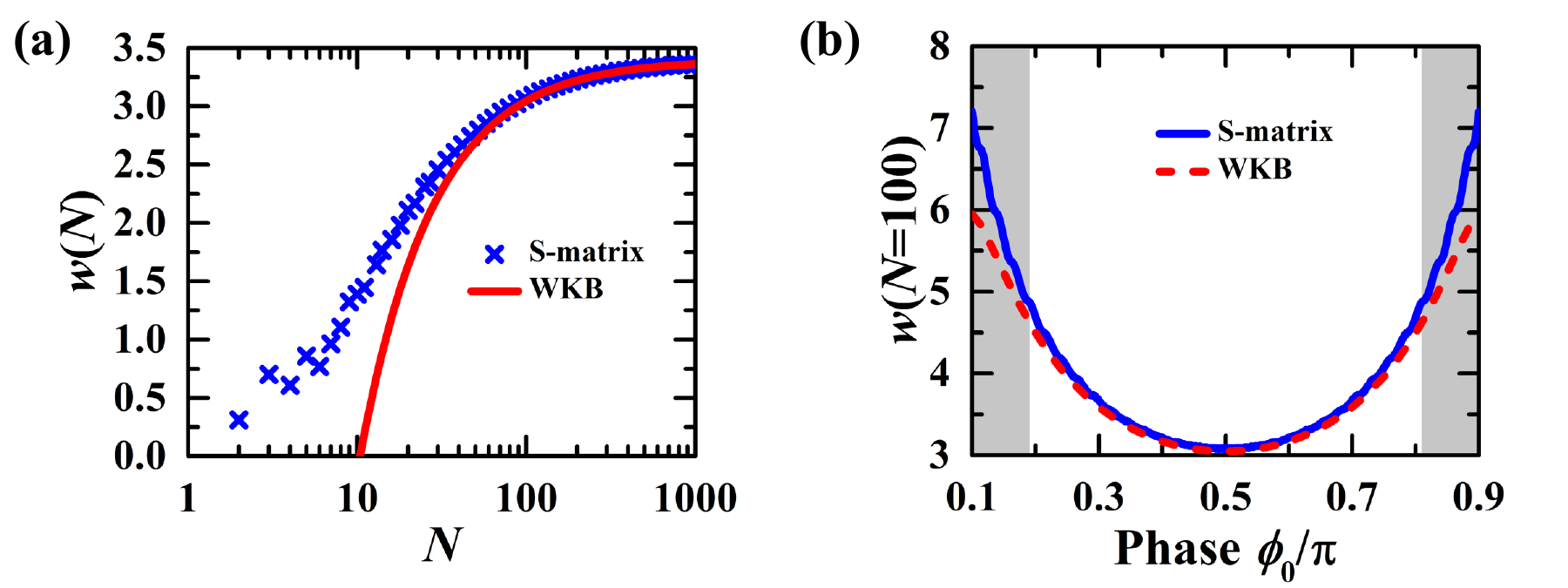} 
\par\end{centering}
\caption{High-reflectivity bandwidth $w$ (defined by $|r|^{2}>0.99$). Discrete
markers are scattering-matrix numerical results; solid curves are
the semiclassical prediction of Eqs.~(\ref{eq:scaling})--(\ref{eq:lambda}).
(a) $w$ versus emitter number $N$ at the anti-Bragg condition ($\phi_{0}=0.5\pi$),
showing the $1/N$ contraction toward the bulk limit and its breakdown
at small $N$, where too few emitters lie within the envelope for
a local band edge to exist. (b) $w$ versus interatomic phase $\phi_{0}$
for $N=100$, $\eta=0.5$, $\sigma=0.1$. The bandwidth is minimal
at the anti-Bragg point and grows on approach to either Bragg point.
The semiclassical curve is drawn only over the interval in which Eq.~(\ref{eq:criterion})
is satisfied. Outside it (shaded) the finite-size correction is no
longer small and the expansion carries no predictive weight. }
\label{fig:WKB} 
\end{figure}

We benchmark the scaling relation against exact $S$-matrix results.
In Fig.~\ref{fig:WKB}(a) we plot the reflection bandwidth $w$ versus
emitter number $N$ at the anti-Bragg condition $\phi_{0}=0.5\pi$.
As $N$ grows, the boundary-induced Fabry-Perot oscillations weaken
and the bandwidth converges toward the bulk asymptote $w_{\infty}$,
which the analytic function $w_{\infty}+\Lambda/N$ (solid curve)
reproduces to within about $1\%$ for $N\gtrsim80$. The residual
discrepancy originates from the parabolic approximation to the barrier
rather than from the semiclassical treatment itself, since evaluating
the attenuation phase of Eq.~(\ref{eq:theta_def}) numerically with
the exact envelope removes most of it.

At small $N$, the scaling law breaks down. At $\eta=0.5$ and $\sigma=0.1$,
the prediction falls $2\%$ below the exact bandwidth at $N=60$,
$5\%$ below at $N=40$ and $22\%$ below at $N=20$. This breakdown
is consistent with Eq.~(\ref{eq:criterion}), since $\sigma N\min(\phi_{0},\pi-\phi_{0})$
equals $9.4$, $6.3$ and $3.1$ at these three sizes, and the finite-size
correction is no longer small. The semiclassical picture assumes a
local band edge, which requires an array that is locally periodic
over the tunneling region. At $N=20$, however, only three emitters
carry more than half the peak rate. Such a small number of strongly
coupled emitters does not support a well-defined band structure, and
the concepts of a barrier and of evanescent leakage through it no
longer apply.

This limitation is relevant here, because the array studied in Fig.~\ref{fig:fig02},
which is also representative of near-term experiments, has $N=20$.
In this regime, the scaling law should be regarded as an extrapolation
rather than a prediction, and the enhancement is instead explained
by the spectral factorization of Sec.~\ref{sec:spectral}, which
does not rely on a band edge and remains exact at any $N$. The pole
analysis accounts for the enhancement in short arrays, while the scaling
law provides its closed-form dependence on $N$, $\eta$, $\sigma$
and $\phi_{0}$ once the array supports a local band edge.

Figure~\ref{fig:WKB}(b) shows the bandwidth as a function of the
interatomic phase $\phi_{0}$ at fixed $N=100$. The window is narrowest
at the anti-Bragg point and widens monotonically toward either Bragg
point, following the growth of the bulk bandgap $\gamma_{\text{max}}/\sin\phi_{0}$.
Over the central range, the semiclassical curve agrees with the exact
result within a few percent, and the agreement degrades toward either
Bragg point, as predicted by Eq.~(\ref{eq:criterion}). The deviation
grows from $0.7\%$ at $\phi_{0}=\pi/2$ to $3.8\%$ at both $\phi_{0}=0.2\pi$
and $0.8\pi$, and to $17\%$ at both $\phi_{0}=0.1\pi$ and $0.9\pi$.
At these three phases, $\sigma N\min(\phi_{0},\pi-\phi_{0})$ equals
$15.7$, $6.3$ and $3.1$. This deviation reflects the breakdown
of the expansion rather than a change in the physics. The exact bandwidth
continues to grow toward the Bragg points, where a uniform array acts
as an ideal mirror, and only the truncated scaling law loses its accuracy.
We restrict the analytic curve in Fig.~\ref{fig:WKB}(b) to $\sigma N\min(\phi_{0},\pi-\phi_{0})\gtrsim6$.

\section{Discussion and conclusion}

\label{sec:conclusion}

In this work, we have investigated single-photon reflection from a
waveguide QED array with a Gaussian transverse-displacement profile
and the resulting exponential-of-Gaussian emitter-waveguide coupling
profile. Our results show that, at a fixed total decay rate, the profile
broadens the coherent reflection band and suppresses the boundary-induced
Fabry-Perot oscillations. The spectral factorization attributes both
effects to the collective poles. At non-Bragg spacings, the transmission
zeros remain pinned at the atomic resonance, while the spatial modulation
redistributes the pole widths and prevents an individual subradiant
mode from terminating the reflection window. We further examined the
finite-size behavior using a semiclassical WKB treatment. For sufficiently
large arrays, the bandwidth follows the $1/N$ scaling law in Eqs.~(\ref{eq:scaling})--(\ref{eq:lambda}).
The bulk term is determined by the peak local decay rate $\gamma_{\text{max}}$,
whereas the finite-size correction grows as the envelope narrows.
The competition between these terms selects an optimal profile width.

We note three limitations of the present analysis. First, the scaling
law is controlled by the product of $\sigma N$ with the distance
to the nearer Bragg point rather than by $N$ alone, so its accuracy
decreases toward either Bragg point. For the parameters used in Fig.~\ref{fig:fig02},
this product is $\approx3$, and the analytic result should be regarded
as an extrapolation rather than a controlled prediction. In this regime,
the exact bandwidth still increases toward the Bragg points. Second,
we restrict the analysis to the single-excitation sector, so the nonlinear
response at higher photon number remains to be addressed. Finally,
the Markovian description assumes negligible retardation across the
array, an assumption that must be revisited for arrays long enough
that the propagation delay becomes comparable to the collective decay
time.

The proposed protocol is compatible with current experimental platforms.
It requires site-dependent control of the guided-mode decay rates
while the emitter spacing and transition frequencies remain approximately
fixed. For neutral atoms near optical nanofibers or photonic-crystal
waveguides, such control can be achieved through transverse positioning
in the evanescent field~\citep{LeKien2004,Goban2012,Goban2014}.
In circuit-QED arrays, the coupling profile can be adjusted through
the geometry of the capacitive coupling to an open transmission line~\citep{GarciaRipoll2015}.
A quantitative implementation should also account for fabrication
constraints and decay into nonguided channels, which are not included
in the present lossless model.

Looking ahead, several promising directions remain. A systematic non-Markovian
treatment is needed to determine how retardation modifies the finite-size
scaling and the degeneracy of the transmission zeros~\citep{Andersson2019}.
Extending the analysis beyond the single-excitation sector would reveal
whether the broadband response survives saturation at higher photon
number. In addition, combining the coupling profile with selective-radiance
protocols may connect bandwidth engineering to photon storage in ordered
arrays~\citep{AsenjoGarcia2017,Manzoni2018}. These findings provide
insights into how spatial coupling profiles control collective resonances
and offer practical design principles for broadband atomic mirrors.

\section*{Acknowledgments}

We acknowledge support from the National Science and Technology Council
(NSTC), Taiwan, under the Grants No. 112-2112-M-001-079-MY3, NSTC-115-2112-M-001035-MY3
and No. NSTC-115-2119-M-001-009, and from Academia Sinica under Grant
AS-CDA-113-M04. We are also grateful for support from TG 1.2 of NCTS
at NTU.

\appendix
%dummy comment inserted by tex2lyx to ensure that this paragraph is not empty%dummy comment inserted by tex2lyx to ensure that this paragraph is not empty%dummy comment inserted by tex2lyx to ensure that this paragraph is not empty%dummy comment inserted by tex2lyx to ensure that this paragraph is not empty
\global\long\def\appendixname{APPENDIX}%

\section{Matrix determinant ratios for transport coefficients}

\label{app:det}

In this appendix we derive the determinant representations of the
transmission amplitude $t$ and the reflection amplitude $r$ used
in Sec.~\ref{sec:spectral}, and establish the exact degeneracy of
the transmission zeros. Mapping the input-output relations onto exact
matrix identities links the transport coefficients directly to the
roots of two polynomials.

\subsection{Matrix determinant lemma for transmission ($t$)}

For the forward channel we invoke the matrix determinant lemma. For
any invertible $\mathbf{A}\in\mathbb{C}^{N\times N}$ and column vectors
$\mathbf{u},\mathbf{v}\in\mathbb{C}^{N\times1}$, the determinant
of the rank-one perturbation satisfies 
\begin{equation}
\det(\mathbf{A}+\mathbf{u}\mathbf{v}^{\dagger})=\det(\mathbf{A})\left(1+\mathbf{v}^{\dagger}\mathbf{A}^{-1}\mathbf{u}\right).\label{eq:mdl}
\end{equation}
In the present system we identify $\mathbf{A}=G^{-1}=\omega_{\text{in}}I-H_{\text{eff}}$
and assign the perturbations $\mathbf{u}=-i\mathbf{u}_{R}$ and $\mathbf{v}=\mathbf{u}_{R}$,
with $\mathbf{u}_{R}$ the right-incoming channel vector. Defining
the modified forward-scattering matrix 
\begin{equation}
A_{t}\equiv G^{-1}-i\mathbf{u}_{R}\mathbf{u}_{R}^{\dagger},\label{eq:At}
\end{equation}
Eq.~(\ref{eq:mdl}) gives $\det(A_{t})=\det(G^{-1})\left(1-i\mathbf{u}_{R}^{\dagger}G\mathbf{u}_{R}\right)$,
in which the bracket is the transmission amplitude of Eq.~(\ref{eq:tr}).
Hence 
\begin{equation}
\det\left(A_{t}\right)=\det\left(G^{-1}\right)\,t,\qquad t=\frac{\det(A_{t})}{\det(G^{-1})}.\label{eq:t_ratio}
\end{equation}

\subsection{Nilpotency of $A_{t}$ and the degenerate numerator zeros}

\label{app:nilpotent}

The matrix $A_{t}$ has a structure that determines its spectrum completely.
Using Eq.~(\ref{eq:Ginv}) and $(\mathbf{u}_{R})_{j}=\frac{1}{\sqrt{2}}\sqrt{\gamma_{j}}e^{i\phi_{0}j}$,
the off-diagonal part of Eq.~(\ref{eq:At}) reads 
\begin{equation}
\left(A_{t}-\Delta\,I\right)_{jl}=\frac{i}{2}\sqrt{\gamma_{j}\gamma_{l}}\left[e^{i\phi_{0}|j-l|}-e^{i\phi_{0}(j-l)}\right],\label{eq:At_offdiag}
\end{equation}
with $\Delta$ the probe detuning of Sec.~\ref{sec:model}, where
we have used $\left(\mathbf{u}_{R}\mathbf{u}_{R}^{\dagger}\right)_{jl}=\frac{1}{2}\sqrt{\gamma_{j}\gamma_{l}}e^{i\phi_{0}(j-l)}$.
The bracket vanishes identically for $j\geq l$, since $|j-l|=j-l$
there, so $A_{t}-\Delta\,I$ is strictly upper triangular and therefore
nilpotent. Its determinant is the product of its (vanishing) diagonal
entries shifted by $\Delta$, giving Eq.~(\ref{eq:zeros_exact}),
\[
\det\left(A_{t}\right)=\Delta^{N}=(\omega_{\text{in}}-\omega_{0})^{N}.
\]
This algebraic identity holds for arbitrary positive $\{\gamma_{j}\}$,
arbitrary $N$, and arbitrary $\phi_{0}$ within the present reciprocal,
lossless Markovian model. It is a property of the kernel $e^{i\phi_{0}|j-l|}$,
and it gives $t=(\omega_{\text{in}}-\omega_{0})^{N}/\det(G^{-1})$.
The roots of the numerator correspond to physical transmission zeros
only after common factors with $\det(G^{-1})$ are removed. At the
exact Bragg phases, $N-1$ dark-state poles lie at $\omega_{0}$ and
cancel $N-1$ of these roots, as shown in Eq.~(\ref{eq:bragg_cancellation});
at the non-Bragg phases studied in the main text, no such cancellation
occurs.

Since $A_{t}-\Delta\,I$ is nilpotent, its eigenvalues are highly
sensitive to numerical round-off, and a direct diagonalization of
$A_{t}$ does not reliably resolve the degenerate numerator zeros.
We therefore obtain them from Eq.~(\ref{eq:zeros_exact}), which
places all $N$ algebraic roots at $\omega_{0}$ before any common
factors with $\det(G^{-1})$ are removed.

\subsection{Schur complement theorem for reflection ($r$)}

For the backward channel we use the Schur complement theorem for block
matrices. If the top-left block $\mathbf{A}$ of a partitioned square
matrix is invertible, the determinant factorizes as 
\begin{equation}
\det\begin{pmatrix}\mathbf{A} & \mathbf{B}\\
\mathbf{C} & \mathbf{D}
\end{pmatrix}=\det(\mathbf{A})\cdot\det\left(\mathbf{D}-\mathbf{C}\mathbf{A}^{-1}\mathbf{B}\right).\label{eq:schur}
\end{equation}
We construct the augmented matrix $A_{r}\in\mathbb{C}^{(N+1)\times(N+1)}$
by assigning $\mathbf{A}=G^{-1}$, $\mathbf{B}=\mathbf{u}_{R}$, $\mathbf{C}=i\mathbf{u}_{L}^{\dagger}$,
and $\mathbf{D}=0$, where $\mathbf{u}_{L}$ is the left-outgoing
channel vector. Its determinant follows from Eq.~(\ref{eq:schur}),
\[
\det\left(A_{r}\right)=\det\begin{pmatrix}G^{-1} & \mathbf{u}_{R}\\
i\mathbf{u}_{L}^{\dagger} & 0
\end{pmatrix}=\det\left(G^{-1}\right)\det\left(-i\mathbf{u}_{L}^{\dagger}G\mathbf{u}_{R}\right).
\]
The Schur complement here is a scalar, so its determinant is trivial
and, identifying the reflection amplitude of Eq.~(\ref{eq:tr}),
\begin{equation}
\det\left(A_{r}\right)=\det\left(G^{-1}\right)\,r,\qquad r=\frac{\det(A_{r})}{\det(G^{-1})}.\label{eq:r_ratio}
\end{equation}
Note that $\det(A_{r})=-i\,\mathbf{u}_{L}^{\dagger}\text{adj}(G^{-1})\mathbf{u}_{R}$
is a polynomial of degree $N-1$ in $\omega_{\text{in}}$, one order
lower than $\det(A_{t})$, consistent with $r\to0$ and $t\to1$ at
large detuning. Both transport amplitudes are thus determined by the
polynomial spectra of their respective non-Hermitian operators. The
numerator of $t$ is profile independent by Eq.~(\ref{eq:zeros_exact}),
whereas $\det(A_{r})=-i\mathbf{u}_{L}^{\dagger}\operatorname{adj}(G^{-1})\mathbf{u}_{R}$
retains an explicit dependence on the coupling profile. In the lossless
model, the reflection probability nevertheless follows from $|r|^{2}=1-|t|^{2}$;
the complex reflection amplitude and its phase retain the additional
numerator dependence.

\section{Analytic derivation of the reflection bandwidth scaling law}

\label{app:wkb}

In this appendix we derive the finite-size scaling function $w(N)=w_{\infty}+\Lambda/N$
and obtain explicit forms for the bulk bandwidth $w_{\infty}$ and
the finite-size amplitude $\Lambda$. Section~\ref{app:wkb-a} sets
up the continuum picture and shows why the correction is algebraic
in $1/N$. The quantitative coefficients are then obtained in Secs.~\ref{app:wkb-b}
and \ref{app:wkb-c} from the exact unit-cell transfer matrix, which
remains valid at arbitrary $\phi_{0}$.

\subsection{Continuum wave equation}

\label{app:wkb-a}

We begin by connecting the microscopic lattice dynamics to the macroscopic
transport response. As stated in the main text, eliminating the photonic
reservoir reduces the stationary response of the array to an effective
non-Hermitian Hamiltonian $H_{\text{eff}}$. To obtain the collective
band structure independently of the external drive, we consider the
eigenmodes of $H_{\text{eff}}$ in the single-excitation subspace,
which satisfy the homogeneous lattice equation 
\begin{equation}
-i(\omega-\omega_{0})b_{j}=-\frac{1}{2}\sum_{l=1}^{N}\sqrt{\gamma_{j}\gamma_{l}}\,e^{ik_{0}|z_{j}-z_{l}|}b_{l},\label{eq:lattice}
\end{equation}
where $b_{j}$ is the stationary amplitude of the $j$-th emitter,
$\omega_{0}$ the bare transition frequency, $\gamma_{j}$ the engineered
local decay rate, and $k_{0}=\phi_{0}/d$ the resonant wavevector
expressed through the single-spacing phase $\phi_{0}$.

Under the slowly varying envelope approximation, in which the profile
changes little over many lattice sites ($\sigma N\gg1$), the array
is locally uniform and admits a continuum description. Mapping the
discrete coordinate $z_{j}=j\,d$ onto a continuous variable $z$
converts the sum into an integral over the array length $L=Nd$, $\sum_{l}\rightarrow\int_{0}^{L}dz'/d$.
Introducing the auxiliary wavefunction $f(z)=\sqrt{\gamma(z)}\,b(z)$---where
$b(z)$ is the continuum limit of the discrete amplitudes $b_{j}$
and $\gamma(z)$ that of the envelope defined in Eq.~(\ref{eq:gamma_profile})---removes
the explicit $\sqrt{\gamma_{j}\gamma_{l}}$ prefactor and gives the
integral equation 
\begin{equation}
-i\left[\frac{\omega-\omega_{0}}{\gamma(z)}\right]f(z)=-\frac{1}{2d}\int_{0}^{L}dz'\,e^{ik_{0}\left|z-z'\right|}f(z').\label{eq:integral_eq}
\end{equation}

The absolute-value kernel is eliminated by differentiating twice with
respect to $z$ using Leibniz's rule. Neglecting the gradients of
the slow envelope, which allows $\gamma(z)$ to be treated as locally
constant, yields a Helmholtz-like equation, 
\begin{equation}
\frac{d^{2}f(z)}{dz^{2}}+q^{2}(z)f(z)=0,\qquad q^{2}(z)=k_{0}^{2}-\frac{k_{0}}{d}\frac{\gamma(z)}{\omega-\omega_{0}}.\label{eq:helmholtz}
\end{equation}

The local band edge $\omega_{c}(z)$ is the classical turning point
at which the wave changes from propagating to evanescent, $q^{2}(z)=0$,
giving $\omega_{c}(z)=\omega_{0}+\gamma(z)/\phi_{0}$. This continuum
estimate reduces to the band edge $\omega_{0}+[\gamma(z)/2]\cot(\phi_{0}/2)$
obtained from the unit-cell transfer matrix in Sec.~\ref{app:wkb-b}
for $\phi_{0}\ll1$. Equation~(\ref{eq:helmholtz}) therefore provides
the qualitative picture of tunneling through a barrier, while the
quantitative coefficients are obtained from the transfer matrix. Inside
the gap the wavevector is purely imaginary ($q^{2}<0$), and it is
convenient to define the real attenuation coefficient $\kappa_{z}^{2}(z)=-q^{2}(z)$.

Expanding $\omega_{c}(z)$ to second order about the array center
($\xi=z/L=1/2$), where the profile peaks at $\gamma_{\text{max}}$,
gives a parabolic barrier, 
\begin{equation}
\omega_{c}(\xi)\approx\omega_{\text{max}}-\beta\left(\xi-\tfrac{1}{2}\right)^{2},\label{eq:parabolic}
\end{equation}
with $\omega_{\text{max}}=\omega_{0}+\gamma_{\text{max}}/\phi_{0}$
and curvature $\beta=\gamma_{\text{max}}/(2\sigma_{\text{eff}}^{2}\phi_{0})$,
in which $\sigma_{\text{eff}}=\sigma/(2\sqrt{\eta})$ is the width
of $\gamma(\xi)$ itself, Eq.~(\ref{eq:sigma_eff}), and not that
of the Gaussian appearing in its exponent.

Near the turning points, where $\kappa_{z}=0$, a linear approximation
to the barrier maps Eq.~(\ref{eq:helmholtz}) onto Airy equations~\citep{Bender1999}.
Matching the asymptotic solutions across the turning points gives
the Kemble transmission probability~\citep{Kemble1935}, $T=\left(1+e^{2\theta}\right)^{-1}$,
which in the high-reflectivity regime ($T\ll1$) reduces at leading
order to $T(\omega)\approx e^{-2\theta}$, with 
\begin{equation}
\theta(\omega,N)=\int_{z_{1}}^{z_{2}}\kappa_{z}(z)\,dz\label{eq:theta_def}
\end{equation}
the attenuation phase accumulated between the two turning points $z_{1,2}$,
at which $\kappa_{z}$ vanishes. For the parabolic barrier of Eq.~(\ref{eq:parabolic})
the turning points are $\xi_{1,2}=1/2\pm\sqrt{\Delta\omega/\beta}$
with $\Delta\omega=\omega_{\text{max}}-\omega$, and the integral
evaluates to a form linear in $\Delta\omega$, $\theta=KN\Delta\omega$,
with $K$ a constant fixed by $\phi_{0}$, $\sigma_{\text{eff}}$
and $\gamma_{\text{max}}$. Imposing a fixed transmission threshold
$T(\omega_{\text{edge}})=T_{0},$ with $T_{0}=1-\left|r\right|_{\text{th}}^{2}$
therefore fixes the band-edge contraction as $\Delta\omega=-\ln T_{0}/(KN)$,
algebraic in $1/N$. Subtracting this contraction from the bulk bandgap
at each of the two band edges gives Eq.~(\ref{eq:scaling}). The
coefficient $K$ itself is obtained from the transfer matrix in Sec.~\ref{app:wkb-c}.

\subsection{Bulk bandwidth limit: $w_{\infty}$}

\label{app:wkb-b}

As $N\to\infty$ the finite-size correction vanishes and the bandwidth
approaches its bulk value $w(\infty)=w_{\infty}$. By Bloch's theorem
for an infinite periodic medium~\citep{Yariv2003,Deutsch1995}, the
boundaries between the allowed band and the evanescent gap occur where
the eigenvalues of the unit-cell transfer matrix 
\begin{equation}
\tilde{M}=M_{\text{prop}}M_{\text{atom}}=\begin{pmatrix}e^{i\phi_{0}} & 0\\
0 & e^{-i\phi_{0}}
\end{pmatrix}\begin{pmatrix}1+2\chi & 2\chi\\
-2\chi & 1-2\chi
\end{pmatrix},\label{eq:transfer}
\end{equation}
with $\chi=-i\gamma(z)/\left[4(\omega-\omega_{0})\right]$ the local
atomic polarizability, become degenerate on the unit circle, which
is equivalent to the trace condition $\text{Tr}(\tilde{M})=\pm2$.
The transfer matrix is the product of a free-propagation layer and
a scattering layer describing the local atomic polarizability. Its
trace is 
\begin{equation}
\text{Tr}(\tilde{M})=2\cos\phi_{0}+\frac{\gamma(z)}{\omega-\omega_{0}}\sin\phi_{0}.\label{eq:trace}
\end{equation}
Solving this condition separately for $\text{Tr}(\tilde{M})=+2$ and
$\text{Tr}(\tilde{M})=-2$ gives the two band-edge detunings, 
\begin{equation}
\delta\omega_{\text{bulk},+}=\frac{\gamma(z)}{2}\cot\frac{\phi_{0}}{2},\qquad\delta\omega_{\text{bulk},-}=-\frac{\gamma(z)}{2}\tan\frac{\phi_{0}}{2},\label{eq:edges}
\end{equation}
measured from $\omega_{0}$. The local gap width is their difference,
$w_{\text{loc}}(z)=\gamma(z)/\sin\phi_{0}$. In the bulk limit the
band edge is set by the peak of the envelope, $\gamma(z)\to\gamma_{\text{max}}$,
giving Eq.~(\ref{eq:winf}), 
\[
w_{\infty}(\eta,\sigma,\phi_{0})=\frac{\gamma_{\text{max}}}{\sin\phi_{0}}.
\]

\subsection{Finite-size correction: $\Lambda$}

\label{app:wkb-c}

The finite-size contraction arises from evanescent tunneling near
both band edges, whose local positions $\delta\omega_{\text{bulk},\pm}(z)\propto\gamma(z)$
follow the modulation. To quantify this leakage for a slowly varying
envelope we expand the trace dispersion relation near threshold. Inside
the gap the Bloch phase per unit cell is imaginary, $\cosh(\kappa_{z}d)=|\text{Tr}\,\tilde{M}|/2$,
so that at leading order 
\begin{equation}
(\kappa_{z}d)^{2}\approx\left|\text{Tr}(\tilde{M})\right|-2.\label{eq:kz2}
\end{equation}

Near the upper band edge, shifting the frequency below the bulk limit
by $\Delta\omega_{+}$ and expanding the profile quadratically about
the array center as $\gamma(z)\approx\gamma_{\text{max}}\left[1-(z-L/2)^{2}/2\sigma_{\text{eff}}^{2}L^{2}\right]$
{[}Eq.~(\ref{eq:sigma_eff}){]}, Eqs.~(\ref{eq:trace})--(\ref{eq:kz2})
give the local evanescent wavevector 
\begin{equation}
\kappa_{z}(z)=\frac{2}{d}\sin\frac{\phi_{0}}{2}\sqrt{\frac{\Delta\omega_{+}}{\delta\omega_{\text{bulk},+}}-\frac{(z-L/2)^{2}}{2\sigma_{\text{eff}}^{2}L^{2}}}.\label{eq:kz_upper}
\end{equation}
The wave is evanescent between the turning points $z=L/2\pm z_{\text{turn}}$,
with $z_{\text{turn}}=\sigma_{\text{eff}}L\sqrt{2\Delta\omega_{+}/\delta\omega_{\text{bulk},+}}$,
and integrating across them gives the upper tunneling phase 
\begin{equation}
\theta_{+}=\int_{L/2-z_{\text{turn}}}^{L/2+z_{\text{turn}}}\kappa_{z}(z)\,dz=2\sqrt{2}\,\pi\sigma_{\text{eff}}N\frac{\sin^{2}(\phi_{0}/2)}{\cos(\phi_{0}/2)}\frac{\Delta\omega_{+}}{\gamma_{\text{max}}}.\label{eq:theta_plus}
\end{equation}
Matching the transmission $T=e^{-2\theta_{+}}$ to the threshold condition
$2\theta_{+}=-\ln\left(1-|r|_{\text{th}}^{2}\right)$ isolates the
upper band-edge shrinkage, 
\begin{equation}
\Delta\omega_{+}=\frac{-\ln(1-|r|_{\text{th}}^{2})\,\gamma_{\text{max}}}{4\sqrt{2}\,\pi\sigma_{\text{eff}}N}\frac{\cos(\phi_{0}/2)}{\sin^{2}(\phi_{0}/2)}.\label{eq:dw_plus}
\end{equation}
The same expansion at the lower band edge gives $\theta_{-}=2\sqrt{2}\,\pi\sigma_{\text{eff}}N\left[\cos^{2}(\phi_{0}/2)/\sin(\phi_{0}/2)\right]\Delta\omega_{-}/\gamma_{\text{max}}$
and, under the same criterion, 
\begin{equation}
\Delta\omega_{-}=\frac{-\ln(1-|r|_{\text{th}}^{2})\,\gamma_{\text{max}}}{4\sqrt{2}\,\pi\sigma_{\text{eff}}N}\frac{\sin(\phi_{0}/2)}{\cos^{2}(\phi_{0}/2)}.\label{eq:dw_minus}
\end{equation}
The two band edges move inward independently, so the bandwidth is
\begin{equation}
w(N)=w_{\infty}-\left(\Delta\omega_{+}+\Delta\omega_{-}\right),\label{eq:combine}
\end{equation}
and since $\Delta\omega_{\pm}\propto1/N$ this is precisely Eq.~(\ref{eq:scaling})
with 
\[
\Lambda=\frac{\gamma_{\text{max}}\ln(1-|r|_{\text{th}}^{2})}{4\sqrt{2}\,\pi\sigma_{\text{eff}}}\left[\frac{\cos(\phi_{0}/2)}{\sin^{2}(\phi_{0}/2)}+\frac{\sin(\phi_{0}/2)}{\cos^{2}(\phi_{0}/2)}\right],
\]
which is negative because $|r|_{\text{th}}^{2}<1$. At the anti-Bragg
point the bracket equals $2\sqrt{2}$ and the expression collapses
to $\Lambda=\gamma_{\text{max}}\sqrt{\eta}\ln(1-|r|_{\text{th}}^{2})/(\pi\sigma)$.

\bibliography{main}

\begin{thebibliography}{46}%
\makeatletter
\providecommand \@ifxundefined [1]{%
 \@ifx{#1\undefined}
}%
\providecommand \@ifnum [1]{%
 \ifnum #1\expandafter \@firstoftwo
 \else \expandafter \@secondoftwo
 \fi
}%
\providecommand \@ifx [1]{%
 \ifx #1\expandafter \@firstoftwo
 \else \expandafter \@secondoftwo
 \fi
}%
\providecommand \natexlab [1]{#1}%
\providecommand \enquote  [1]{``#1''}%
\providecommand \bibnamefont  [1]{#1}%
\providecommand \bibfnamefont [1]{#1}%
\providecommand \citenamefont [1]{#1}%
\providecommand \href@noop [0]{\@secondoftwo}%
\providecommand \href [0]{\begingroup \@sanitize@url \@href}%
\providecommand \@href[1]{\@@startlink{#1}\@@href}%
\providecommand \@@href[1]{\endgroup#1\@@endlink}%
\providecommand \@sanitize@url [0]{\catcode `\\12\catcode `\$12\catcode
  `\&12\catcode `\#12\catcode `\^12\catcode `\_12\catcode `\%12\relax}%
\providecommand \@@startlink[1]{}%
\providecommand \@@endlink[0]{}%
\providecommand \url  [0]{\begingroup\@sanitize@url \@url }%
\providecommand \@url [1]{\endgroup\@href {#1}{\urlprefix }}%
\providecommand \urlprefix  [0]{URL }%
\providecommand \Eprint [0]{\href }%
\providecommand \doibase [0]{https://doi.org/}%
\providecommand \selectlanguage [0]{\@gobble}%
\providecommand \bibinfo  [0]{\@secondoftwo}%
\providecommand \bibfield  [0]{\@secondoftwo}%
\providecommand \translation [1]{[#1]}%
\providecommand \BibitemOpen [0]{}%
\providecommand \bibitemStop [0]{}%
\providecommand \bibitemNoStop [0]{.\EOS\space}%
\providecommand \EOS [0]{\spacefactor3000\relax}%
\providecommand \BibitemShut  [1]{\csname bibitem#1\endcsname}%
\let\auto@bib@innerbib\@empty
%</preamble>
\bibitem [{\citenamefont {Shen}\ and\ \citenamefont {Fan}(2005)}]{Shen2005}%
  \BibitemOpen
  \bibfield  {author} {\bibinfo {author} {\bibfnamefont {J.-T.}\ \bibnamefont
  {Shen}}\ and\ \bibinfo {author} {\bibfnamefont {S.}~\bibnamefont {Fan}},\
  }\bibfield  {title} {\bibinfo {title} {Coherent single photon transport in a
  one-dimensional waveguide coupled with superconducting quantum bits},\ }\href
  {https://doi.org/10.1103/physrevlett.95.213001} {\bibfield  {journal}
  {\bibinfo  {journal} {Physical Review Letters}\ }\textbf {\bibinfo {volume}
  {95}},\ \bibinfo {pages} {213001} (\bibinfo {year} {2005})}\BibitemShut
  {NoStop}%
\bibitem [{\citenamefont {Zhou}\ \emph {et~al.}(2008)\citenamefont {Zhou},
  \citenamefont {Dong}, \citenamefont {Liu}, \citenamefont {Sun},\ and\
  \citenamefont {Nori}}]{Zhou2008}%
  \BibitemOpen
  \bibfield  {author} {\bibinfo {author} {\bibfnamefont {L.}~\bibnamefont
  {Zhou}}, \bibinfo {author} {\bibfnamefont {H.}~\bibnamefont {Dong}}, \bibinfo
  {author} {\bibfnamefont {Y.-x.}\ \bibnamefont {Liu}}, \bibinfo {author}
  {\bibfnamefont {C.~P.}\ \bibnamefont {Sun}},\ and\ \bibinfo {author}
  {\bibfnamefont {F.}~\bibnamefont {Nori}},\ }\bibfield  {title} {\bibinfo
  {title} {Quantum supercavity with atomic mirrors},\ }\href
  {https://doi.org/10.1103/physreva.78.063827} {\bibfield  {journal} {\bibinfo
  {journal} {Physical Review A}\ }\textbf {\bibinfo {volume} {78}},\ \bibinfo
  {pages} {063827} (\bibinfo {year} {2008})}\BibitemShut {NoStop}%
\bibitem [{\citenamefont {Roy}\ \emph {et~al.}(2017)\citenamefont {Roy},
  \citenamefont {Wilson},\ and\ \citenamefont {Firstenberg}}]{Roy2017}%
  \BibitemOpen
  \bibfield  {author} {\bibinfo {author} {\bibfnamefont {D.}~\bibnamefont
  {Roy}}, \bibinfo {author} {\bibfnamefont {C.~M.}\ \bibnamefont {Wilson}},\
  and\ \bibinfo {author} {\bibfnamefont {O.}~\bibnamefont {Firstenberg}},\
  }\bibfield  {title} {\bibinfo {title} {Colloquium : Strongly interacting
  photons in one-dimensional continuum},\ }\href
  {https://doi.org/10.1103/revmodphys.89.021001} {\bibfield  {journal}
  {\bibinfo  {journal} {Reviews of Modern Physics}\ }\textbf {\bibinfo {volume}
  {89}},\ \bibinfo {pages} {021001} (\bibinfo {year} {2017})}\BibitemShut
  {NoStop}%
\bibitem [{\citenamefont {Chang}\ \emph {et~al.}(2018)\citenamefont {Chang},
  \citenamefont {Douglas}, \citenamefont {Gonz{\'a}lez-Tudela}, \citenamefont
  {Hung},\ and\ \citenamefont {Kimble}}]{Chang2018}%
  \BibitemOpen
  \bibfield  {author} {\bibinfo {author} {\bibfnamefont {D.~E.}\ \bibnamefont
  {Chang}}, \bibinfo {author} {\bibfnamefont {J.~S.}\ \bibnamefont {Douglas}},
  \bibinfo {author} {\bibfnamefont {A.}~\bibnamefont {Gonz{\'a}lez-Tudela}},
  \bibinfo {author} {\bibfnamefont {C.-L.}\ \bibnamefont {Hung}},\ and\
  \bibinfo {author} {\bibfnamefont {H.~J.}\ \bibnamefont {Kimble}},\ }\bibfield
   {title} {\bibinfo {title} {Colloquium: Quantum matter built from nanoscopic
  lattices of atoms and photons},\ }\href
  {https://doi.org/10.1103/revmodphys.90.031002} {\bibfield  {journal}
  {\bibinfo  {journal} {Reviews of Modern Physics}\ }\textbf {\bibinfo {volume}
  {90}},\ \bibinfo {pages} {031002} (\bibinfo {year} {2018})}\BibitemShut
  {NoStop}%
\bibitem [{\citenamefont {Sheremet}\ \emph {et~al.}(2023)\citenamefont
  {Sheremet}, \citenamefont {Petrov}, \citenamefont {Iorsh}, \citenamefont
  {Poshakinskiy},\ and\ \citenamefont {Poddubny}}]{Sheremet2023}%
  \BibitemOpen
  \bibfield  {author} {\bibinfo {author} {\bibfnamefont {A.~S.}\ \bibnamefont
  {Sheremet}}, \bibinfo {author} {\bibfnamefont {M.~I.}\ \bibnamefont
  {Petrov}}, \bibinfo {author} {\bibfnamefont {I.~V.}\ \bibnamefont {Iorsh}},
  \bibinfo {author} {\bibfnamefont {A.~V.}\ \bibnamefont {Poshakinskiy}},\ and\
  \bibinfo {author} {\bibfnamefont {A.~N.}\ \bibnamefont {Poddubny}},\
  }\bibfield  {title} {\bibinfo {title} {Waveguide quantum electrodynamics:
  Collective radiance and photon-photon correlations},\ }\href
  {https://doi.org/10.1103/revmodphys.95.015002} {\bibfield  {journal}
  {\bibinfo  {journal} {Reviews of Modern Physics}\ }\textbf {\bibinfo {volume}
  {95}},\ \bibinfo {pages} {015002} (\bibinfo {year} {2023})}\BibitemShut
  {NoStop}%
\bibitem [{\citenamefont {Lalumi{\`e}re}\ \emph {et~al.}(2013)\citenamefont
  {Lalumi{\`e}re}, \citenamefont {Sanders}, \citenamefont {van Loo},
  \citenamefont {Fedorov}, \citenamefont {Wallraff},\ and\ \citenamefont
  {Blais}}]{Lalumiere2013}%
  \BibitemOpen
  \bibfield  {author} {\bibinfo {author} {\bibfnamefont {K.}~\bibnamefont
  {Lalumi{\`e}re}}, \bibinfo {author} {\bibfnamefont {B.~C.}\ \bibnamefont
  {Sanders}}, \bibinfo {author} {\bibfnamefont {A.~F.}\ \bibnamefont {van
  Loo}}, \bibinfo {author} {\bibfnamefont {A.}~\bibnamefont {Fedorov}},
  \bibinfo {author} {\bibfnamefont {A.}~\bibnamefont {Wallraff}},\ and\
  \bibinfo {author} {\bibfnamefont {A.}~\bibnamefont {Blais}},\ }\bibfield
  {title} {\bibinfo {title} {Input-output theory for waveguide qed with an
  ensemble of inhomogeneous atoms},\ }\href
  {https://doi.org/10.1103/physreva.88.043806} {\bibfield  {journal} {\bibinfo
  {journal} {Physical Review A}\ }\textbf {\bibinfo {volume} {88}},\ \bibinfo
  {pages} {043806} (\bibinfo {year} {2013})}\BibitemShut {NoStop}%
\bibitem [{\citenamefont {Solano}\ \emph {et~al.}(2017)\citenamefont {Solano},
  \citenamefont {Barberis-Blostein}, \citenamefont {Fatemi}, \citenamefont
  {Orozco},\ and\ \citenamefont {Rolston}}]{Solano2017}%
  \BibitemOpen
  \bibfield  {author} {\bibinfo {author} {\bibfnamefont {P.}~\bibnamefont
  {Solano}}, \bibinfo {author} {\bibfnamefont {P.}~\bibnamefont
  {Barberis-Blostein}}, \bibinfo {author} {\bibfnamefont {F.~K.}\ \bibnamefont
  {Fatemi}}, \bibinfo {author} {\bibfnamefont {L.~A.}\ \bibnamefont {Orozco}},\
  and\ \bibinfo {author} {\bibfnamefont {S.~L.}\ \bibnamefont {Rolston}},\
  }\bibfield  {title} {\bibinfo {title} {Super-radiance reveals infinite-range
  dipole interactions through a nanofiber},\ }\href
  {https://doi.org/10.1038/s41467-017-01994-3} {\bibfield  {journal} {\bibinfo
  {journal} {Nature Communications}\ }\textbf {\bibinfo {volume} {8}},\
  \bibinfo {pages} {1857} (\bibinfo {year} {2017})}\BibitemShut {NoStop}%
\bibitem [{\citenamefont {Astafiev}\ \emph {et~al.}(2010)\citenamefont
  {Astafiev}, \citenamefont {Zagoskin}, \citenamefont {Abdumalikov},
  \citenamefont {Pashkin}, \citenamefont {Yamamoto}, \citenamefont {Inomata},
  \citenamefont {Nakamura},\ and\ \citenamefont {Tsai}}]{Astafiev2010}%
  \BibitemOpen
  \bibfield  {author} {\bibinfo {author} {\bibfnamefont {O.}~\bibnamefont
  {Astafiev}}, \bibinfo {author} {\bibfnamefont {A.~M.}\ \bibnamefont
  {Zagoskin}}, \bibinfo {author} {\bibfnamefont {A.~A.}\ \bibnamefont
  {Abdumalikov}}, \bibinfo {author} {\bibfnamefont {Y.~A.}\ \bibnamefont
  {Pashkin}}, \bibinfo {author} {\bibfnamefont {T.}~\bibnamefont {Yamamoto}},
  \bibinfo {author} {\bibfnamefont {K.}~\bibnamefont {Inomata}}, \bibinfo
  {author} {\bibfnamefont {Y.}~\bibnamefont {Nakamura}},\ and\ \bibinfo
  {author} {\bibfnamefont {J.~S.}\ \bibnamefont {Tsai}},\ }\bibfield  {title}
  {\bibinfo {title} {Resonance fluorescence of a single artificial atom},\
  }\href {https://doi.org/10.1126/science.1181918} {\bibfield  {journal}
  {\bibinfo  {journal} {Science}\ }\textbf {\bibinfo {volume} {327}},\ \bibinfo
  {pages} {840} (\bibinfo {year} {2010})}\BibitemShut {NoStop}%
\bibitem [{\citenamefont {Hoi}\ \emph {et~al.}(2011)\citenamefont {Hoi},
  \citenamefont {Wilson}, \citenamefont {Johansson}, \citenamefont {Palomaki},
  \citenamefont {Peropadre},\ and\ \citenamefont {Delsing}}]{Hoi2011}%
  \BibitemOpen
  \bibfield  {author} {\bibinfo {author} {\bibfnamefont {I.-C.}\ \bibnamefont
  {Hoi}}, \bibinfo {author} {\bibfnamefont {C.~M.}\ \bibnamefont {Wilson}},
  \bibinfo {author} {\bibfnamefont {G.}~\bibnamefont {Johansson}}, \bibinfo
  {author} {\bibfnamefont {T.}~\bibnamefont {Palomaki}}, \bibinfo {author}
  {\bibfnamefont {B.}~\bibnamefont {Peropadre}},\ and\ \bibinfo {author}
  {\bibfnamefont {P.}~\bibnamefont {Delsing}},\ }\bibfield  {title} {\bibinfo
  {title} {Demonstration of a single-photon router in the microwave regime},\
  }\href {https://doi.org/10.1103/physrevlett.107.073601} {\bibfield  {journal}
  {\bibinfo  {journal} {Physical Review Letters}\ }\textbf {\bibinfo {volume}
  {107}},\ \bibinfo {pages} {073601} (\bibinfo {year} {2011})}\BibitemShut
  {NoStop}%
\bibitem [{\citenamefont {van Loo}\ \emph {et~al.}(2013)\citenamefont {van
  Loo}, \citenamefont {Fedorov}, \citenamefont {Lalumi{\`e}re}, \citenamefont
  {Sanders}, \citenamefont {Blais},\ and\ \citenamefont
  {Wallraff}}]{vanLoo2013}%
  \BibitemOpen
  \bibfield  {author} {\bibinfo {author} {\bibfnamefont {A.~F.}\ \bibnamefont
  {van Loo}}, \bibinfo {author} {\bibfnamefont {A.}~\bibnamefont {Fedorov}},
  \bibinfo {author} {\bibfnamefont {K.}~\bibnamefont {Lalumi{\`e}re}}, \bibinfo
  {author} {\bibfnamefont {B.~C.}\ \bibnamefont {Sanders}}, \bibinfo {author}
  {\bibfnamefont {A.}~\bibnamefont {Blais}},\ and\ \bibinfo {author}
  {\bibfnamefont {A.}~\bibnamefont {Wallraff}},\ }\bibfield  {title} {\bibinfo
  {title} {Photon-mediated interactions between distant artificial atoms},\
  }\href {https://doi.org/10.1126/science.1244324} {\bibfield  {journal}
  {\bibinfo  {journal} {Science}\ }\textbf {\bibinfo {volume} {342}},\ \bibinfo
  {pages} {1494} (\bibinfo {year} {2013})}\BibitemShut {NoStop}%
\bibitem [{\citenamefont {Mirhosseini}\ \emph {et~al.}(2019)\citenamefont
  {Mirhosseini}, \citenamefont {Kim}, \citenamefont {Zhang}, \citenamefont
  {Sipahigil}, \citenamefont {Dieterle}, \citenamefont {Keller}, \citenamefont
  {Asenjo-Garcia}, \citenamefont {Chang},\ and\ \citenamefont
  {Painter}}]{Mirhosseini2019}%
  \BibitemOpen
  \bibfield  {author} {\bibinfo {author} {\bibfnamefont {M.}~\bibnamefont
  {Mirhosseini}}, \bibinfo {author} {\bibfnamefont {E.}~\bibnamefont {Kim}},
  \bibinfo {author} {\bibfnamefont {X.}~\bibnamefont {Zhang}}, \bibinfo
  {author} {\bibfnamefont {A.}~\bibnamefont {Sipahigil}}, \bibinfo {author}
  {\bibfnamefont {P.~B.}\ \bibnamefont {Dieterle}}, \bibinfo {author}
  {\bibfnamefont {A.~J.}\ \bibnamefont {Keller}}, \bibinfo {author}
  {\bibfnamefont {A.}~\bibnamefont {Asenjo-Garcia}}, \bibinfo {author}
  {\bibfnamefont {D.~E.}\ \bibnamefont {Chang}},\ and\ \bibinfo {author}
  {\bibfnamefont {O.}~\bibnamefont {Painter}},\ }\bibfield  {title} {\bibinfo
  {title} {Cavity quantum electrodynamics with atom-like mirrors},\ }\href
  {https://doi.org/10.1038/s41586-019-1196-1} {\bibfield  {journal} {\bibinfo
  {journal} {Nature}\ }\textbf {\bibinfo {volume} {569}},\ \bibinfo {pages}
  {692} (\bibinfo {year} {2019})}\BibitemShut {NoStop}%
\bibitem [{\citenamefont {Vetsch}\ \emph {et~al.}(2010)\citenamefont {Vetsch},
  \citenamefont {Reitz}, \citenamefont {Sagu{\'e}}, \citenamefont {Schmidt},
  \citenamefont {Dawkins},\ and\ \citenamefont {Rauschenbeutel}}]{Vetsch2010}%
  \BibitemOpen
  \bibfield  {author} {\bibinfo {author} {\bibfnamefont {E.}~\bibnamefont
  {Vetsch}}, \bibinfo {author} {\bibfnamefont {D.}~\bibnamefont {Reitz}},
  \bibinfo {author} {\bibfnamefont {G.}~\bibnamefont {Sagu{\'e}}}, \bibinfo
  {author} {\bibfnamefont {R.}~\bibnamefont {Schmidt}}, \bibinfo {author}
  {\bibfnamefont {S.~T.}\ \bibnamefont {Dawkins}},\ and\ \bibinfo {author}
  {\bibfnamefont {A.}~\bibnamefont {Rauschenbeutel}},\ }\bibfield  {title}
  {\bibinfo {title} {Optical interface created by laser-cooled atoms trapped in
  the evanescent field surrounding an optical nanofiber},\ }\href
  {https://doi.org/10.1103/physrevlett.104.203603} {\bibfield  {journal}
  {\bibinfo  {journal} {Physical Review Letters}\ }\textbf {\bibinfo {volume}
  {104}},\ \bibinfo {pages} {203603} (\bibinfo {year} {2010})}\BibitemShut
  {NoStop}%
\bibitem [{\citenamefont {Goban}\ \emph {et~al.}(2012)\citenamefont {Goban},
  \citenamefont {Choi}, \citenamefont {Alton}, \citenamefont {Ding},
  \citenamefont {Lacro{\^u}te}, \citenamefont {Pototschnig}, \citenamefont
  {Thiele}, \citenamefont {Stern},\ and\ \citenamefont {Kimble}}]{Goban2012}%
  \BibitemOpen
  \bibfield  {author} {\bibinfo {author} {\bibfnamefont {A.}~\bibnamefont
  {Goban}}, \bibinfo {author} {\bibfnamefont {K.~S.}\ \bibnamefont {Choi}},
  \bibinfo {author} {\bibfnamefont {D.~J.}\ \bibnamefont {Alton}}, \bibinfo
  {author} {\bibfnamefont {D.}~\bibnamefont {Ding}}, \bibinfo {author}
  {\bibfnamefont {C.}~\bibnamefont {Lacro{\^u}te}}, \bibinfo {author}
  {\bibfnamefont {M.}~\bibnamefont {Pototschnig}}, \bibinfo {author}
  {\bibfnamefont {T.}~\bibnamefont {Thiele}}, \bibinfo {author} {\bibfnamefont
  {N.~P.}\ \bibnamefont {Stern}},\ and\ \bibinfo {author} {\bibfnamefont
  {H.~J.}\ \bibnamefont {Kimble}},\ }\bibfield  {title} {\bibinfo {title}
  {Demonstration of a state-insensitive, compensated nanofiber trap},\ }\href
  {https://doi.org/10.1103/physrevlett.109.033603} {\bibfield  {journal}
  {\bibinfo  {journal} {Physical Review Letters}\ }\textbf {\bibinfo {volume}
  {109}},\ \bibinfo {pages} {033603} (\bibinfo {year} {2012})}\BibitemShut
  {NoStop}%
\bibitem [{\citenamefont {Corzo}\ \emph {et~al.}(2019)\citenamefont {Corzo},
  \citenamefont {Raskop}, \citenamefont {Chandra}, \citenamefont {Sheremet},
  \citenamefont {Gouraud},\ and\ \citenamefont {Laurat}}]{Corzo2019}%
  \BibitemOpen
  \bibfield  {author} {\bibinfo {author} {\bibfnamefont {N.~V.}\ \bibnamefont
  {Corzo}}, \bibinfo {author} {\bibfnamefont {J.}~\bibnamefont {Raskop}},
  \bibinfo {author} {\bibfnamefont {A.}~\bibnamefont {Chandra}}, \bibinfo
  {author} {\bibfnamefont {A.~S.}\ \bibnamefont {Sheremet}}, \bibinfo {author}
  {\bibfnamefont {B.}~\bibnamefont {Gouraud}},\ and\ \bibinfo {author}
  {\bibfnamefont {J.}~\bibnamefont {Laurat}},\ }\bibfield  {title} {\bibinfo
  {title} {Waveguide-coupled single collective excitation of atomic arrays},\
  }\href {https://doi.org/10.1038/s41586-019-0902-3} {\bibfield  {journal}
  {\bibinfo  {journal} {Nature}\ }\textbf {\bibinfo {volume} {566}},\ \bibinfo
  {pages} {359} (\bibinfo {year} {2019})}\BibitemShut {NoStop}%
\bibitem [{\citenamefont {Goban}\ \emph {et~al.}(2014)\citenamefont {Goban},
  \citenamefont {Hung}, \citenamefont {Yu}, \citenamefont {Hood}, \citenamefont
  {Muniz}, \citenamefont {Lee}, \citenamefont {Martin}, \citenamefont
  {McClung}, \citenamefont {Choi}, \citenamefont {Chang}, \citenamefont
  {Painter},\ and\ \citenamefont {Kimble}}]{Goban2014}%
  \BibitemOpen
  \bibfield  {author} {\bibinfo {author} {\bibfnamefont {A.}~\bibnamefont
  {Goban}}, \bibinfo {author} {\bibfnamefont {C.-L.}\ \bibnamefont {Hung}},
  \bibinfo {author} {\bibfnamefont {S.-P.}\ \bibnamefont {Yu}}, \bibinfo
  {author} {\bibfnamefont {J.~D.}\ \bibnamefont {Hood}}, \bibinfo {author}
  {\bibfnamefont {J.~A.}\ \bibnamefont {Muniz}}, \bibinfo {author}
  {\bibfnamefont {J.~H.}\ \bibnamefont {Lee}}, \bibinfo {author} {\bibfnamefont
  {M.~J.}\ \bibnamefont {Martin}}, \bibinfo {author} {\bibfnamefont {A.~C.}\
  \bibnamefont {McClung}}, \bibinfo {author} {\bibfnamefont {K.~S.}\
  \bibnamefont {Choi}}, \bibinfo {author} {\bibfnamefont {D.~E.}\ \bibnamefont
  {Chang}}, \bibinfo {author} {\bibfnamefont {O.}~\bibnamefont {Painter}},\
  and\ \bibinfo {author} {\bibfnamefont {H.~J.}\ \bibnamefont {Kimble}},\
  }\bibfield  {title} {\bibinfo {title} {Atom-light interactions in photonic
  crystals},\ }\href {https://doi.org/10.1038/ncomms4808} {\bibfield  {journal}
  {\bibinfo  {journal} {Nature Communications}\ }\textbf {\bibinfo {volume}
  {5}},\ \bibinfo {pages} {3808} (\bibinfo {year} {2014})}\BibitemShut
  {NoStop}%
\bibitem [{\citenamefont {Goban}\ \emph {et~al.}(2015)\citenamefont {Goban},
  \citenamefont {Hung}, \citenamefont {Hood}, \citenamefont {Yu}, \citenamefont
  {Muniz}, \citenamefont {Painter},\ and\ \citenamefont {Kimble}}]{Goban2015}%
  \BibitemOpen
  \bibfield  {author} {\bibinfo {author} {\bibfnamefont {A.}~\bibnamefont
  {Goban}}, \bibinfo {author} {\bibfnamefont {C.-L.}\ \bibnamefont {Hung}},
  \bibinfo {author} {\bibfnamefont {J.~D.}\ \bibnamefont {Hood}}, \bibinfo
  {author} {\bibfnamefont {S.-P.}\ \bibnamefont {Yu}}, \bibinfo {author}
  {\bibfnamefont {J.~A.}\ \bibnamefont {Muniz}}, \bibinfo {author}
  {\bibfnamefont {O.}~\bibnamefont {Painter}},\ and\ \bibinfo {author}
  {\bibfnamefont {H.~J.}\ \bibnamefont {Kimble}},\ }\bibfield  {title}
  {\bibinfo {title} {Superradiance for atoms trapped along a photonic crystal
  waveguide},\ }\href {https://doi.org/10.1103/physrevlett.115.063601}
  {\bibfield  {journal} {\bibinfo  {journal} {Physical Review Letters}\
  }\textbf {\bibinfo {volume} {115}},\ \bibinfo {pages} {063601} (\bibinfo
  {year} {2015})}\BibitemShut {NoStop}%
\bibitem [{\citenamefont {Deutsch}\ \emph {et~al.}(1995)\citenamefont
  {Deutsch}, \citenamefont {Spreeuw}, \citenamefont {Rolston},\ and\
  \citenamefont {Phillips}}]{Deutsch1995}%
  \BibitemOpen
  \bibfield  {author} {\bibinfo {author} {\bibfnamefont {I.~H.}\ \bibnamefont
  {Deutsch}}, \bibinfo {author} {\bibfnamefont {R.~J.~C.}\ \bibnamefont
  {Spreeuw}}, \bibinfo {author} {\bibfnamefont {S.~L.}\ \bibnamefont
  {Rolston}},\ and\ \bibinfo {author} {\bibfnamefont {W.~D.}\ \bibnamefont
  {Phillips}},\ }\bibfield  {title} {\bibinfo {title} {Photonic band gaps in
  optical lattices},\ }\href {https://doi.org/10.1103/physreva.52.1394}
  {\bibfield  {journal} {\bibinfo  {journal} {Physical Review A}\ }\textbf
  {\bibinfo {volume} {52}},\ \bibinfo {pages} {1394} (\bibinfo {year}
  {1995})}\BibitemShut {NoStop}%
\bibitem [{\citenamefont {Chang}\ \emph {et~al.}(2012)\citenamefont {Chang},
  \citenamefont {Jiang}, \citenamefont {Gorshkov},\ and\ \citenamefont
  {Kimble}}]{Chang2012}%
  \BibitemOpen
  \bibfield  {author} {\bibinfo {author} {\bibfnamefont {D.~E.}\ \bibnamefont
  {Chang}}, \bibinfo {author} {\bibfnamefont {L.}~\bibnamefont {Jiang}},
  \bibinfo {author} {\bibfnamefont {A.~V.}\ \bibnamefont {Gorshkov}},\ and\
  \bibinfo {author} {\bibfnamefont {H.~J.}\ \bibnamefont {Kimble}},\ }\bibfield
   {title} {\bibinfo {title} {Cavity qed with atomic mirrors},\ }\href
  {https://doi.org/10.1088/1367-2630/14/6/063003} {\bibfield  {journal}
  {\bibinfo  {journal} {New Journal of Physics}\ }\textbf {\bibinfo {volume}
  {14}},\ \bibinfo {pages} {063003} (\bibinfo {year} {2012})}\BibitemShut
  {NoStop}%
\bibitem [{\citenamefont {Corzo}\ \emph {et~al.}(2016)\citenamefont {Corzo},
  \citenamefont {Gouraud}, \citenamefont {Chandra}, \citenamefont {Goban},
  \citenamefont {Sheremet}, \citenamefont {Kupriyanov},\ and\ \citenamefont
  {Laurat}}]{Corzo2016}%
  \BibitemOpen
  \bibfield  {author} {\bibinfo {author} {\bibfnamefont {N.~V.}\ \bibnamefont
  {Corzo}}, \bibinfo {author} {\bibfnamefont {B.}~\bibnamefont {Gouraud}},
  \bibinfo {author} {\bibfnamefont {A.}~\bibnamefont {Chandra}}, \bibinfo
  {author} {\bibfnamefont {A.}~\bibnamefont {Goban}}, \bibinfo {author}
  {\bibfnamefont {A.~S.}\ \bibnamefont {Sheremet}}, \bibinfo {author}
  {\bibfnamefont {D.~V.}\ \bibnamefont {Kupriyanov}},\ and\ \bibinfo {author}
  {\bibfnamefont {J.}~\bibnamefont {Laurat}},\ }\bibfield  {title} {\bibinfo
  {title} {Large bragg reflection from one-dimensional chains of trapped atoms
  near a nanoscale waveguide},\ }\href
  {https://doi.org/10.1103/physrevlett.117.133603} {\bibfield  {journal}
  {\bibinfo  {journal} {Physical Review Letters}\ }\textbf {\bibinfo {volume}
  {117}},\ \bibinfo {pages} {133603} (\bibinfo {year} {2016})}\BibitemShut
  {NoStop}%
\bibitem [{\citenamefont {S{\o}rensen}\ \emph {et~al.}(2016)\citenamefont
  {S{\o}rensen}, \citenamefont {B{\'e}guin}, \citenamefont {Kluge},
  \citenamefont {Iakoupov}, \citenamefont {S{\o}rensen}, \citenamefont
  {M{\"u}ller}, \citenamefont {Polzik},\ and\ \citenamefont
  {Appel}}]{Sorensen2016}%
  \BibitemOpen
  \bibfield  {author} {\bibinfo {author} {\bibfnamefont {H.~L.}\ \bibnamefont
  {S{\o}rensen}}, \bibinfo {author} {\bibfnamefont {J.-B.}\ \bibnamefont
  {B{\'e}guin}}, \bibinfo {author} {\bibfnamefont {K.~W.}\ \bibnamefont
  {Kluge}}, \bibinfo {author} {\bibfnamefont {I.}~\bibnamefont {Iakoupov}},
  \bibinfo {author} {\bibfnamefont {A.~S.}\ \bibnamefont {S{\o}rensen}},
  \bibinfo {author} {\bibfnamefont {J.~H.}\ \bibnamefont {M{\"u}ller}},
  \bibinfo {author} {\bibfnamefont {E.~S.}\ \bibnamefont {Polzik}},\ and\
  \bibinfo {author} {\bibfnamefont {J.}~\bibnamefont {Appel}},\ }\bibfield
  {title} {\bibinfo {title} {Coherent backscattering of light off
  one-dimensional atomic strings},\ }\href
  {https://doi.org/10.1103/physrevlett.117.133604} {\bibfield  {journal}
  {\bibinfo  {journal} {Physical Review Letters}\ }\textbf {\bibinfo {volume}
  {117}},\ \bibinfo {pages} {133604} (\bibinfo {year} {2016})}\BibitemShut
  {NoStop}%
\bibitem [{\citenamefont {Rui}\ \emph {et~al.}(2020)\citenamefont {Rui},
  \citenamefont {Wei}, \citenamefont {Rubio-Abadal}, \citenamefont {Hollerith},
  \citenamefont {Zeiher}, \citenamefont {Stamper-Kurn}, \citenamefont {Gross},\
  and\ \citenamefont {Bloch}}]{Rui2020}%
  \BibitemOpen
  \bibfield  {author} {\bibinfo {author} {\bibfnamefont {J.}~\bibnamefont
  {Rui}}, \bibinfo {author} {\bibfnamefont {D.}~\bibnamefont {Wei}}, \bibinfo
  {author} {\bibfnamefont {A.}~\bibnamefont {Rubio-Abadal}}, \bibinfo {author}
  {\bibfnamefont {S.}~\bibnamefont {Hollerith}}, \bibinfo {author}
  {\bibfnamefont {J.}~\bibnamefont {Zeiher}}, \bibinfo {author} {\bibfnamefont
  {D.~M.}\ \bibnamefont {Stamper-Kurn}}, \bibinfo {author} {\bibfnamefont
  {C.}~\bibnamefont {Gross}},\ and\ \bibinfo {author} {\bibfnamefont
  {I.}~\bibnamefont {Bloch}},\ }\bibfield  {title} {\bibinfo {title} {A
  subradiant optical mirror formed by a single structured atomic layer},\
  }\href {https://doi.org/10.1038/s41586-020-2463-x} {\bibfield  {journal}
  {\bibinfo  {journal} {Nature}\ }\textbf {\bibinfo {volume} {583}},\ \bibinfo
  {pages} {369} (\bibinfo {year} {2020})}\BibitemShut {NoStop}%
\bibitem [{\citenamefont {Shahmoon}\ \emph {et~al.}(2017)\citenamefont
  {Shahmoon}, \citenamefont {Wild}, \citenamefont {Lukin},\ and\ \citenamefont
  {Yelin}}]{Shahmoon2017}%
  \BibitemOpen
  \bibfield  {author} {\bibinfo {author} {\bibfnamefont {E.}~\bibnamefont
  {Shahmoon}}, \bibinfo {author} {\bibfnamefont {D.~S.}\ \bibnamefont {Wild}},
  \bibinfo {author} {\bibfnamefont {M.~D.}\ \bibnamefont {Lukin}},\ and\
  \bibinfo {author} {\bibfnamefont {S.~F.}\ \bibnamefont {Yelin}},\ }\bibfield
  {title} {\bibinfo {title} {Cooperative resonances in light scattering from
  two-dimensional atomic arrays},\ }\href
  {https://doi.org/10.1103/physrevlett.118.113601} {\bibfield  {journal}
  {\bibinfo  {journal} {Physical Review Letters}\ }\textbf {\bibinfo {volume}
  {118}},\ \bibinfo {pages} {113601} (\bibinfo {year} {2017})}\BibitemShut
  {NoStop}%
\bibitem [{\citenamefont {Yariv}(2003)}]{Yariv2003}%
  \BibitemOpen
  \bibfield  {author} {\bibinfo {author} {\bibfnamefont {A.}~\bibnamefont
  {Yariv}},\ }\href@noop {} {\emph {\bibinfo {title} {Optical waves in
  crystals}}},\ \bibinfo {edition} {wiley classics library ed}\ ed.,\ edited
  by\ \bibinfo {editor} {\bibfnamefont {P.}~\bibnamefont {Yeh}},\ Wiley
  classics library\ (\bibinfo  {publisher} {Wiley},\ \bibinfo {address}
  {Hoboken, NJ},\ \bibinfo {year} {2003})\ \bibinfo {note}
  {"Wiley-Interscience." - Originally published: New York, c1984 - Includes
  bibliographies and index}\BibitemShut {NoStop}%
\bibitem [{\citenamefont {Poddubny}(2022)}]{Poddubny2022}%
  \BibitemOpen
  \bibfield  {author} {\bibinfo {author} {\bibfnamefont {A.~N.}\ \bibnamefont
  {Poddubny}},\ }\bibfield  {title} {\bibinfo {title} {Driven anti-bragg
  subradiant correlations in waveguide quantum electrodynamics},\ }\href
  {https://doi.org/10.1103/physreva.106.l031702} {\bibfield  {journal}
  {\bibinfo  {journal} {Physical Review A}\ }\textbf {\bibinfo {volume}
  {106}},\ \bibinfo {pages} {L031702} (\bibinfo {year} {2022})}\BibitemShut
  {NoStop}%
\bibitem [{\citenamefont {Zhang}\ \emph {et~al.}(2026)\citenamefont {Zhang},
  \citenamefont {Ke}, \citenamefont {Peng}, \citenamefont {Chen}, \citenamefont
  {Liu}, \citenamefont {Zhang},\ and\ \citenamefont {Lee}}]{Zhang2026}%
  \BibitemOpen
  \bibfield  {author} {\bibinfo {author} {\bibfnamefont {X.}~\bibnamefont
  {Zhang}}, \bibinfo {author} {\bibfnamefont {Y.}~\bibnamefont {Ke}}, \bibinfo
  {author} {\bibfnamefont {Z.}~\bibnamefont {Peng}}, \bibinfo {author}
  {\bibfnamefont {Z.}~\bibnamefont {Chen}}, \bibinfo {author} {\bibfnamefont
  {W.}~\bibnamefont {Liu}}, \bibinfo {author} {\bibfnamefont {L.}~\bibnamefont
  {Zhang}},\ and\ \bibinfo {author} {\bibfnamefont {C.}~\bibnamefont {Lee}},\
  }\bibfield  {title} {\bibinfo {title} {Localization and scattering of a
  photon in quasiperiodic qubit arrays},\ }\href
  {https://doi.org/10.1103/2mct-vck8} {\bibfield  {journal} {\bibinfo
  {journal} {Physical Review A}\ }\textbf {\bibinfo {volume} {113}},\ \bibinfo
  {pages} {063704} (\bibinfo {year} {2026})}\BibitemShut {NoStop}%
\bibitem [{\citenamefont {Asenjo-Garcia}\ \emph {et~al.}(2017)\citenamefont
  {Asenjo-Garcia}, \citenamefont {Moreno-Cardoner}, \citenamefont {Albrecht},
  \citenamefont {Kimble},\ and\ \citenamefont {Chang}}]{AsenjoGarcia2017}%
  \BibitemOpen
  \bibfield  {author} {\bibinfo {author} {\bibfnamefont {A.}~\bibnamefont
  {Asenjo-Garcia}}, \bibinfo {author} {\bibfnamefont {M.}~\bibnamefont
  {Moreno-Cardoner}}, \bibinfo {author} {\bibfnamefont {A.}~\bibnamefont
  {Albrecht}}, \bibinfo {author} {\bibfnamefont {H.~J.}\ \bibnamefont
  {Kimble}},\ and\ \bibinfo {author} {\bibfnamefont {D.~E.}\ \bibnamefont
  {Chang}},\ }\bibfield  {title} {\bibinfo {title} {Exponential improvement in
  photon storage fidelities using subradiance and ``selective radiance'' in
  atomic arrays},\ }\href {https://doi.org/10.1103/physrevx.7.031024}
  {\bibfield  {journal} {\bibinfo  {journal} {Physical Review X}\ }\textbf
  {\bibinfo {volume} {7}},\ \bibinfo {pages} {031024} (\bibinfo {year}
  {2017})}\BibitemShut {NoStop}%
\bibitem [{\citenamefont {Manzoni}\ \emph {et~al.}(2018)\citenamefont
  {Manzoni}, \citenamefont {Moreno-Cardoner}, \citenamefont {Asenjo-Garcia},
  \citenamefont {Porto}, \citenamefont {Gorshkov},\ and\ \citenamefont
  {Chang}}]{Manzoni2018}%
  \BibitemOpen
  \bibfield  {author} {\bibinfo {author} {\bibfnamefont {M.~T.}\ \bibnamefont
  {Manzoni}}, \bibinfo {author} {\bibfnamefont {M.}~\bibnamefont
  {Moreno-Cardoner}}, \bibinfo {author} {\bibfnamefont {A.}~\bibnamefont
  {Asenjo-Garcia}}, \bibinfo {author} {\bibfnamefont {J.~V.}\ \bibnamefont
  {Porto}}, \bibinfo {author} {\bibfnamefont {A.~V.}\ \bibnamefont
  {Gorshkov}},\ and\ \bibinfo {author} {\bibfnamefont {D.~E.}\ \bibnamefont
  {Chang}},\ }\bibfield  {title} {\bibinfo {title} {Optimization of photon
  storage fidelity in ordered atomic arrays},\ }\href
  {https://doi.org/10.1088/1367-2630/aadb74} {\bibfield  {journal} {\bibinfo
  {journal} {New Journal of Physics}\ }\textbf {\bibinfo {volume} {20}},\
  \bibinfo {pages} {083048} (\bibinfo {year} {2018})}\BibitemShut {NoStop}%
\bibitem [{\citenamefont {Jen}(2020)}]{Jen2020}%
  \BibitemOpen
  \bibfield  {author} {\bibinfo {author} {\bibfnamefont {H.~H.}\ \bibnamefont
  {Jen}},\ }\bibfield  {title} {\bibinfo {title} {Disorder-assisted excitation
  localization in chirally coupled quantum emitters},\ }\href
  {https://doi.org/10.1103/physreva.102.043525} {\bibfield  {journal} {\bibinfo
   {journal} {Physical Review A}\ }\textbf {\bibinfo {volume} {102}},\ \bibinfo
  {pages} {043525} (\bibinfo {year} {2020})}\BibitemShut {NoStop}%
\bibitem [{\citenamefont {Jen}(2021)}]{Jen2021}%
  \BibitemOpen
  \bibfield  {author} {\bibinfo {author} {\bibfnamefont {H.~H.}\ \bibnamefont
  {Jen}},\ }\bibfield  {title} {\bibinfo {title} {Bound and subradiant
  multiatom excitations in an atomic array with nonreciprocal couplings},\
  }\href {https://doi.org/10.1103/physreva.103.063711} {\bibfield  {journal}
  {\bibinfo  {journal} {Physical Review A}\ }\textbf {\bibinfo {volume}
  {103}},\ \bibinfo {pages} {063711} (\bibinfo {year} {2021})}\BibitemShut
  {NoStop}%
\bibitem [{\citenamefont {Lee}\ \emph {et~al.}(2025)\citenamefont {Lee},
  \citenamefont {Lin}, \citenamefont {Lin},\ and\ \citenamefont
  {Jen}}]{Lee2025}%
  \BibitemOpen
  \bibfield  {author} {\bibinfo {author} {\bibfnamefont {C.-Y.}\ \bibnamefont
  {Lee}}, \bibinfo {author} {\bibfnamefont {K.-T.}\ \bibnamefont {Lin}},
  \bibinfo {author} {\bibfnamefont {G.-D.}\ \bibnamefont {Lin}},\ and\ \bibinfo
  {author} {\bibfnamefont {H.~H.}\ \bibnamefont {Jen}},\ }\bibfield  {title}
  {\bibinfo {title} {Controlling excitation localization in waveguide qed
  systems},\ }\href {https://doi.org/10.1103/2256-6x36} {\bibfield  {journal}
  {\bibinfo  {journal} {Physical Review Research}\ }\textbf {\bibinfo {volume}
  {7}},\ \bibinfo {pages} {043123} (\bibinfo {year} {2025})}\BibitemShut
  {NoStop}%
\bibitem [{\citenamefont {Wang}\ \emph {et~al.}(2025)\citenamefont {Wang},
  \citenamefont {He}, \citenamefont {Liao},\ and\ \citenamefont
  {Zubairy}}]{Wang2025}%
  \BibitemOpen
  \bibfield  {author} {\bibinfo {author} {\bibfnamefont {X.}~\bibnamefont
  {Wang}}, \bibinfo {author} {\bibfnamefont {J.}~\bibnamefont {He}}, \bibinfo
  {author} {\bibfnamefont {Z.}~\bibnamefont {Liao}},\ and\ \bibinfo {author}
  {\bibfnamefont {M.~S.}\ \bibnamefont {Zubairy}},\ }\bibfield  {title}
  {\bibinfo {title} {Tunable ultrahigh broadband reflection via collective
  atom-atom interaction in a waveguide-qed system},\ }\href
  {https://doi.org/10.1103/physreva.111.013706} {\bibfield  {journal} {\bibinfo
   {journal} {Physical Review A}\ }\textbf {\bibinfo {volume} {111}},\ \bibinfo
  {pages} {013706} (\bibinfo {year} {2025})}\BibitemShut {NoStop}%
\bibitem [{\citenamefont {Sinha}\ \emph {et~al.}(2020)\citenamefont {Sinha},
  \citenamefont {Meystre}, \citenamefont {Goldschmidt}, \citenamefont {Fatemi},
  \citenamefont {Rolston},\ and\ \citenamefont {Solano}}]{Sinha2020}%
  \BibitemOpen
  \bibfield  {author} {\bibinfo {author} {\bibfnamefont {K.}~\bibnamefont
  {Sinha}}, \bibinfo {author} {\bibfnamefont {P.}~\bibnamefont {Meystre}},
  \bibinfo {author} {\bibfnamefont {E.~A.}\ \bibnamefont {Goldschmidt}},
  \bibinfo {author} {\bibfnamefont {F.~K.}\ \bibnamefont {Fatemi}}, \bibinfo
  {author} {\bibfnamefont {S.~L.}\ \bibnamefont {Rolston}},\ and\ \bibinfo
  {author} {\bibfnamefont {P.}~\bibnamefont {Solano}},\ }\bibfield  {title}
  {\bibinfo {title} {Non-markovian collective emission from macroscopically
  separated emitters},\ }\href {https://doi.org/10.1103/physrevlett.124.043603}
  {\bibfield  {journal} {\bibinfo  {journal} {Physical Review Letters}\
  }\textbf {\bibinfo {volume} {124}},\ \bibinfo {pages} {043603} (\bibinfo
  {year} {2020})}\BibitemShut {NoStop}%
\bibitem [{\citenamefont {Zheng}\ and\ \citenamefont
  {Baranger}(2013)}]{Zheng2013}%
  \BibitemOpen
  \bibfield  {author} {\bibinfo {author} {\bibfnamefont {H.}~\bibnamefont
  {Zheng}}\ and\ \bibinfo {author} {\bibfnamefont {H.~U.}\ \bibnamefont
  {Baranger}},\ }\bibfield  {title} {\bibinfo {title} {Persistent quantum beats
  and long-distance entanglement from waveguide-mediated interactions},\ }\href
  {https://doi.org/10.1103/physrevlett.110.113601} {\bibfield  {journal}
  {\bibinfo  {journal} {Physical Review Letters}\ }\textbf {\bibinfo {volume}
  {110}},\ \bibinfo {pages} {113601} (\bibinfo {year} {2013})}\BibitemShut
  {NoStop}%
\bibitem [{\citenamefont {Dinc}\ and\ \citenamefont
  {Brańczyk}(2019)}]{Dinc2019}%
  \BibitemOpen
  \bibfield  {author} {\bibinfo {author} {\bibfnamefont {F.}~\bibnamefont
  {Dinc}}\ and\ \bibinfo {author} {\bibfnamefont {A.~M.}\ \bibnamefont
  {Brańczyk}},\ }\bibfield  {title} {\bibinfo {title} {Non-markovian
  super-superradiance in a linear chain of up to 100 qubits},\ }\href
  {https://doi.org/10.1103/physrevresearch.1.032042} {\bibfield  {journal}
  {\bibinfo  {journal} {Physical Review Research}\ }\textbf {\bibinfo {volume}
  {1}},\ \bibinfo {pages} {032042} (\bibinfo {year} {2019})}\BibitemShut
  {NoStop}%
\bibitem [{\citenamefont {Le~Kien}\ \emph {et~al.}(2004)\citenamefont
  {Le~Kien}, \citenamefont {Balykin},\ and\ \citenamefont
  {Hakuta}}]{LeKien2004}%
  \BibitemOpen
  \bibfield  {author} {\bibinfo {author} {\bibfnamefont {F.}~\bibnamefont
  {Le~Kien}}, \bibinfo {author} {\bibfnamefont {V.~I.}\ \bibnamefont
  {Balykin}},\ and\ \bibinfo {author} {\bibfnamefont {K.}~\bibnamefont
  {Hakuta}},\ }\bibfield  {title} {\bibinfo {title} {Atom trap and waveguide
  using a two-color evanescent light field around a subwavelength-diameter
  optical fiber},\ }\href {https://doi.org/10.1103/physreva.70.063403}
  {\bibfield  {journal} {\bibinfo  {journal} {Physical Review A}\ }\textbf
  {\bibinfo {volume} {70}},\ \bibinfo {pages} {063403} (\bibinfo {year}
  {2004})}\BibitemShut {NoStop}%
\bibitem [{\citenamefont {Gardiner}\ and\ \citenamefont
  {Collett}(1985)}]{Gardiner1985}%
  \BibitemOpen
  \bibfield  {author} {\bibinfo {author} {\bibfnamefont {C.~W.}\ \bibnamefont
  {Gardiner}}\ and\ \bibinfo {author} {\bibfnamefont {M.~J.}\ \bibnamefont
  {Collett}},\ }\bibfield  {title} {\bibinfo {title} {Input and output in
  damped quantum systems: Quantum stochastic differential equations and the
  master equation},\ }\href {https://doi.org/10.1103/physreva.31.3761}
  {\bibfield  {journal} {\bibinfo  {journal} {Physical Review A}\ }\textbf
  {\bibinfo {volume} {31}},\ \bibinfo {pages} {3761} (\bibinfo {year}
  {1985})}\BibitemShut {NoStop}%
\bibitem [{\citenamefont {Moiseyev}(2011)}]{Moiseyev2011}%
  \BibitemOpen
  \bibfield  {author} {\bibinfo {author} {\bibfnamefont {N.}~\bibnamefont
  {Moiseyev}},\ }\href {https://doi.org/10.1017/cbo9780511976186} {\emph
  {\bibinfo {title} {Non-Hermitian Quantum Mechanics}}}\ (\bibinfo  {publisher}
  {Cambridge University Press},\ \bibinfo {year} {2011})\BibitemShut {NoStop}%
\bibitem [{\citenamefont {Ashida}\ \emph {et~al.}(2020)\citenamefont {Ashida},
  \citenamefont {Gong},\ and\ \citenamefont {Ueda}}]{Ashida2020}%
  \BibitemOpen
  \bibfield  {author} {\bibinfo {author} {\bibfnamefont {Y.}~\bibnamefont
  {Ashida}}, \bibinfo {author} {\bibfnamefont {Z.}~\bibnamefont {Gong}},\ and\
  \bibinfo {author} {\bibfnamefont {M.}~\bibnamefont {Ueda}},\ }\bibfield
  {title} {\bibinfo {title} {Non-hermitian physics},\ }\href
  {https://doi.org/10.1080/00018732.2021.1876991} {\bibfield  {journal}
  {\bibinfo  {journal} {Advances in Physics}\ }\textbf {\bibinfo {volume}
  {69}},\ \bibinfo {pages} {249} (\bibinfo {year} {2020})}\BibitemShut
  {NoStop}%
\bibitem [{\citenamefont {Caneva}\ \emph {et~al.}(2015)\citenamefont {Caneva},
  \citenamefont {Manzoni}, \citenamefont {Shi}, \citenamefont {Douglas},
  \citenamefont {Cirac},\ and\ \citenamefont {Chang}}]{Caneva2015}%
  \BibitemOpen
  \bibfield  {author} {\bibinfo {author} {\bibfnamefont {T.}~\bibnamefont
  {Caneva}}, \bibinfo {author} {\bibfnamefont {M.~T.}\ \bibnamefont {Manzoni}},
  \bibinfo {author} {\bibfnamefont {T.}~\bibnamefont {Shi}}, \bibinfo {author}
  {\bibfnamefont {J.~S.}\ \bibnamefont {Douglas}}, \bibinfo {author}
  {\bibfnamefont {J.~I.}\ \bibnamefont {Cirac}},\ and\ \bibinfo {author}
  {\bibfnamefont {D.~E.}\ \bibnamefont {Chang}},\ }\bibfield  {title} {\bibinfo
  {title} {Quantum dynamics of propagating photons with strong interactions: a
  generalized input--output formalism},\ }\href
  {https://doi.org/10.1088/1367-2630/17/11/113001} {\bibfield  {journal}
  {\bibinfo  {journal} {New Journal of Physics}\ }\textbf {\bibinfo {volume}
  {17}},\ \bibinfo {pages} {113001} (\bibinfo {year} {2015})}\BibitemShut
  {NoStop}%
\bibitem [{\citenamefont {Xu}\ and\ \citenamefont {Fan}(2015)}]{Xu2015}%
  \BibitemOpen
  \bibfield  {author} {\bibinfo {author} {\bibfnamefont {S.}~\bibnamefont
  {Xu}}\ and\ \bibinfo {author} {\bibfnamefont {S.}~\bibnamefont {Fan}},\
  }\bibfield  {title} {\bibinfo {title} {Input-output formalism for few-photon
  transport: A systematic treatment beyond two photons},\ }\href
  {https://doi.org/10.1103/physreva.91.043845} {\bibfield  {journal} {\bibinfo
  {journal} {Physical Review A}\ }\textbf {\bibinfo {volume} {91}},\ \bibinfo
  {pages} {043845} (\bibinfo {year} {2015})}\BibitemShut {NoStop}%
\bibitem [{\citenamefont {Shi}\ \emph {et~al.}(2015)\citenamefont {Shi},
  \citenamefont {Chang},\ and\ \citenamefont {Cirac}}]{Shi2015}%
  \BibitemOpen
  \bibfield  {author} {\bibinfo {author} {\bibfnamefont {T.}~\bibnamefont
  {Shi}}, \bibinfo {author} {\bibfnamefont {D.~E.}\ \bibnamefont {Chang}},\
  and\ \bibinfo {author} {\bibfnamefont {J.~I.}\ \bibnamefont {Cirac}},\
  }\bibfield  {title} {\bibinfo {title} {Multiphoton-scattering theory and
  generalized master equations},\ }\href
  {https://doi.org/10.1103/physreva.92.053834} {\bibfield  {journal} {\bibinfo
  {journal} {Physical Review A}\ }\textbf {\bibinfo {volume} {92}},\ \bibinfo
  {pages} {053834} (\bibinfo {year} {2015})}\BibitemShut {NoStop}%
\bibitem [{\citenamefont {Sweeney}\ \emph {et~al.}(2020)\citenamefont
  {Sweeney}, \citenamefont {Hsu},\ and\ \citenamefont {Stone}}]{Sweeney2020}%
  \BibitemOpen
  \bibfield  {author} {\bibinfo {author} {\bibfnamefont {W.~R.}\ \bibnamefont
  {Sweeney}}, \bibinfo {author} {\bibfnamefont {C.~W.}\ \bibnamefont {Hsu}},\
  and\ \bibinfo {author} {\bibfnamefont {A.~D.}\ \bibnamefont {Stone}},\
  }\bibfield  {title} {\bibinfo {title} {Theory of reflectionless scattering
  modes},\ }\href {https://doi.org/10.1103/physreva.102.063511} {\bibfield
  {journal} {\bibinfo  {journal} {Physical Review A}\ }\textbf {\bibinfo
  {volume} {102}},\ \bibinfo {pages} {063511} (\bibinfo {year}
  {2020})}\BibitemShut {NoStop}%
\bibitem [{\citenamefont {Bender}\ and\ \citenamefont
  {Orszag}(1999)}]{Bender1999}%
  \BibitemOpen
  \bibfield  {author} {\bibinfo {author} {\bibfnamefont {C.~M.}\ \bibnamefont
  {Bender}}\ and\ \bibinfo {author} {\bibfnamefont {S.~A.}\ \bibnamefont
  {Orszag}},\ }\href {https://doi.org/10.1007/978-1-4757-3069-2} {\emph
  {\bibinfo {title} {Advanced Mathematical Methods for Scientists and Engineers
  I}}}\ (\bibinfo  {publisher} {Springer New York},\ \bibinfo {year}
  {1999})\BibitemShut {NoStop}%
\bibitem [{\citenamefont {Kemble}(1935)}]{Kemble1935}%
  \BibitemOpen
  \bibfield  {author} {\bibinfo {author} {\bibfnamefont {E.~C.}\ \bibnamefont
  {Kemble}},\ }\bibfield  {title} {\bibinfo {title} {A contribution to the
  theory of the b. w. k. method},\ }\href
  {https://doi.org/10.1103/physrev.48.549} {\bibfield  {journal} {\bibinfo
  {journal} {Physical Review}\ }\textbf {\bibinfo {volume} {48}},\ \bibinfo
  {pages} {549} (\bibinfo {year} {1935})}\BibitemShut {NoStop}%
\bibitem [{\citenamefont {Garc{\'i}a-Ripoll}\ \emph {et~al.}(2015)\citenamefont
  {Garc{\'i}a-Ripoll}, \citenamefont {Peropadre},\ and\ \citenamefont
  {De~Liberato}}]{GarciaRipoll2015}%
  \BibitemOpen
  \bibfield  {author} {\bibinfo {author} {\bibfnamefont {J.~J.}\ \bibnamefont
  {Garc{\'i}a-Ripoll}}, \bibinfo {author} {\bibfnamefont {B.}~\bibnamefont
  {Peropadre}},\ and\ \bibinfo {author} {\bibfnamefont {S.}~\bibnamefont
  {De~Liberato}},\ }\bibfield  {title} {\bibinfo {title} {Light-matter
  decoupling and {$A^2$} term detection in superconducting circuits},\ }\href
  {https://doi.org/10.1038/srep16055} {\bibfield  {journal} {\bibinfo
  {journal} {Scientific Reports}\ }\textbf {\bibinfo {volume} {5}},\ \bibinfo
  {pages} {16055} (\bibinfo {year} {2015})}\BibitemShut {NoStop}%
\bibitem [{\citenamefont {Andersson}\ \emph {et~al.}(2019)\citenamefont
  {Andersson}, \citenamefont {Suri}, \citenamefont {Guo}, \citenamefont
  {Aref},\ and\ \citenamefont {Delsing}}]{Andersson2019}%
  \BibitemOpen
  \bibfield  {author} {\bibinfo {author} {\bibfnamefont {G.}~\bibnamefont
  {Andersson}}, \bibinfo {author} {\bibfnamefont {B.}~\bibnamefont {Suri}},
  \bibinfo {author} {\bibfnamefont {L.}~\bibnamefont {Guo}}, \bibinfo {author}
  {\bibfnamefont {T.}~\bibnamefont {Aref}},\ and\ \bibinfo {author}
  {\bibfnamefont {P.}~\bibnamefont {Delsing}},\ }\bibfield  {title} {\bibinfo
  {title} {Non-exponential decay of a giant artificial atom},\ }\href
  {https://doi.org/10.1038/s41567-019-0605-6} {\bibfield  {journal} {\bibinfo
  {journal} {Nature Physics}\ }\textbf {\bibinfo {volume} {15}},\ \bibinfo
  {pages} {1123} (\bibinfo {year} {2019})}\BibitemShut {NoStop}%
\end{thebibliography}%

\end{document}